\documentclass[12pt]{article}
\usepackage{amssymb}
\usepackage{epsfig}
\catcode`\@=11
\def\@normalsize{
   \@setsize\normalsize{15pt}\xiipt\@xiipt
   \abovedisplayskip 14pt plus3pt minus3pt%
   \belowdisplayskip \abovedisplayskip
   \abovedisplayshortskip  \z@ plus3pt%
   \belowdisplayshortskip  7pt plus3.5pt minus0pt
}
\def\section{\@startsection{section}{1}{\z@}
   {3.5ex plus 1ex minus .2ex}{2.3ex plus .2ex}{\large\bf}
}

\def\ps@headings{
   \def\@oddfoot{}\def\@evenfoot{}
   \def\@oddhead{
      \hbox{}\hfill
      \makebox[.5\textwidth]{\raggedright\ignorespaces --\thepage{}-- \hfill }
   }
   \def\@evenhead{\@oddhead}
   \def\subsectionmark##1{\markboth{##1}{}}
}%
\ps@headings
\def\titlepage{
   \@restonecolfalse\if@twocolumn\@restonecoltrue\onecolumn
   \else \newpage \fi \thispagestyle{empty}\c@page\z@
   \def\thefootnote{\fnsymbol{footnote}}
}
\def\endtitlepage{
   \if@restonecol\twocolumn \else  \fi
   \def\thefootnote{\arabic{footnote}}
   \setcounter{footnote}{0}
}%
\catcode`@=12
\relax
\newcommand{\newc}{\newcommand}

\newc{\ra}{\rightarrow}
\newc{\lra}{\leftrightarrow}
\newc{\beq}{\begin{equation}}
\newc{\be}{\begin{equation}}
\newc{\eeq}{\end{equation}}
\newc{\ee}{\end{equation}}
\newc{\bea}{\begin{eqnarray}}
\newc{\eea}{\end{eqnarray}}
\def\eps{\epsilon}

\newc{\ome}{\omega}
\newc{\ba}{\begin{eqnarray}}
\newc{\ea}{\end{eqnarray}}
\def\tila{\tilde{\alpha}}

\newcommand{\e}{\varepsilon}

\newc{\dsty}{\displaystyle}
\begin{document}
\def\firstpage#1#2#3#4#5#6{
   \begin{titlepage}
      \nopagebreak
      \title{
         \begin{flushright}
         \vspace*{-0.8in}
         {  }
         \end{flushright}
         \vfill {#3}
      }
      \author{\large #4 \\[1.0cm] #5}
      \maketitle
      \vskip -7mm
      \nopagebreak
      \begin{abstract}
         {\noindent #6}
      \end{abstract}
      \vfill
      \begin{flushleft}
         \rule{16.1cm}{0.2mm}\\[-3mm]
      \end{flushleft}
      \thispagestyle{empty}
   \end{titlepage}}
\date{}%
\firstpage{3118}{IC/95/34}%
{\large\bf D-brane Standard Model variants and Split Supersymmetry: Unification
and fermion mass predictions.}%
{D.V.~Gioutsos, G.K.~Leontaris and A.~Psallidas}%
{\normalsize\sl Theoretical Physics Division, Ioannina University, GR-45110
Ioannina, Greece\\[2.5mm]}%
{We study D-brane inspired models with $U(3)\times U(2)\times U(1)^N$ gauge
symmetry in the context of split supersymmetry. We consider  configurations with
one, two and three ($N=1,2,3$) abelian branes and derive  all hypercharge
embeddings which imply a realistic particle content. Then, we analyze the
implications of split supersymmetry  on the magnitude of the string scale, the
gauge coupling evolution, the third family fermion mass relations and the gaugino
masses. We consider  gauge coupling relations which may arise in parallel as well
as intersecting brane scenarios and classify the various models according to
their predictions for the magnitude of the string scale and the low energy
implications. In the parallel brane scenario where the $U(1)$ branes are
superposed to $U(2)$ or $U(3)$ brane stacks, varying the split susy scale in a
wide range, we find three distinct cases of models predicting a high,
intermediate and low string scale, $M_S\sim 10^{16}$ GeV, $M_S\sim 10^{7}$ GeV
and $M_S\sim 10^{4}$ GeV respectively. We further find that in the intermediate
string scale model the low energy ratio $m_b/m_{\tau}$ is compatible with
$b-\tau$ Yukawa unification at the string scale. Furthermore, we perform a
similar analysis for arbitrary abelian gauge coupling relations at $M_S$
corresponding to possible intersecting brane models. We find cases which predict
a string scale of the order $M_S\ge 10^{14}$ GeV that accommodate a right-handed
neutrino mass of the same order so that a see-saw type light left-handed neutrino
component is obtained in the sub-eV range as required by experimental and
cosmological data. Finally, a short discussion is devoted for the gaugino masses
and the life-time of the gluino.}%
\vfill {\it }
\newpage

\section{Introduction}

Over the past few decades, low energy scale Supersymmetry (SUSY) has appeared as
the most efficient mechanism to solve the hierarchy problem and protect the Higgs
mass from unwanted large radiative corrections, in theories which attempt to
unify the gauge couplings of the Standard Model (SM) at a high scale.
Implementing SUSY in these theories, one avoids the fine-tuning of the parameters
to 30 decimals thus, the criterion of  naturalness of the theory is satisfied.
However, there is a much more severe fine-tuning problem,  i.e., that with regard
to the Cosmological Constant which cannot be solved in the context of the
existing theories like supersymmetry, technicolor etc. The solution of this
problem would require new threshold dynamics at a scale as low as $10^{-3}$\,eV.

It has been recently argued  that the invention of a new mechanism which will
solve the Cosmological Constant problem might  offer a solution to the hierarchy
problem as well. Bearing in mind that such a mechanism will also give an
explanation to the hierarchy of the mass scales, recently, the  scenario of split
supersymmetry has been proposed~\cite{Arkani-Hamed:2004fb} according to which
supersymmetry could be broken  at a high scale $\tilde m$ which can be even of
the order of the GUT scale. In this scenario, squarks and sleptons obtain large
masses of the order of the supersymmetry breaking scale $\tilde m$, while the
corresponding fermionic degrees, gauginos and higgsinos, remain light with masses
at the TeV scale. This splitting of the spectrum is possible when the dominant
source of SUSY breaking preserves an R-symmetry which protects fermionic degrees
to obtain masses at the scale $\tilde m$~\cite{Arkani-Hamed:2004yi}. Gauge
coupling unification at a high scale $M_S$ can be achieved, while $b-\tau$ Yukawa
unification at $M_S$ is compatible with the $m_{b}/m_{\tau}$ low energy ratio  in
split supersymmetry~\cite{Giudice:2004tc,Arvanitaki:2004eu}. In has been further
argued that in certain  D-brane constructions, the presence of internal magnetic
fields~\cite{Bachas:1995ik} provide a concrete realization of split
supersymmetry~\cite{Antoniadis:2004dt,Antoniadis:2005em}, therefore, intermediate
or higher string scales are also viable since there is no hierarchy problem in
this case.

In a previous paper~\cite{Gioutsos:2005uw}, based on D-brane constructions
originally proposed in~\cite{Antoniadis:2002en}, we analyzed the gauge coupling
evolution of D-brane inspired models with gauge symmetry $U(3)\times U(2)\times
U(1)^N$ at the string scale. We restricted our analysis in non-supersymmetric
configurations with two or three abelian branes ($N=2,3$) where only one Higgs
doublet couples to the up quarks and a second one to  the down quarks and
leptons, while all fermion and Higgs fields are obtained from strings attached on
different brane stacks. We examined six models which arise in the context of
these two brane configurations due to the different hypercharge embeddings and
different $U(1)$ brane orientations, and calculated the string scale in partial
gauge coupling unification scenarios, where  the $U(1)$ branes are aligned either
to the $U(3)$ or to the $U(2)$ brane stacks. It was shown that the string scale
depends strongly on the particular orientation of the $U(1)$ branes and the
hypercharge embedding. There exist particular embeddings which  allow a string
scale of a few TeV, while in other D-brane configurations the string scale raises
up to intermediate or even higher mass scales. We further investigated the
possibility of obtaining the correct $m_b/m_{\tau}$ relation at $M_Z$  for
$b-\tau$ Yukawa equality at the string scale and found that this is possible for
the class of models with string scale of the order $10^{6-7}$GeV. Further
interesting issues on the effective field theory of this class of
models~\cite{Antoniadis:2002en} have also been analyzed in detail by the authors
of ref.\cite{Coriano:2005js}.

In view of the interesting results obtained in split supersymmetry, as well as
the possibility pointed out in~\cite{Antoniadis:2004dt}, that the D-branes might
provide a natural realization for the spectrum of split SUSY, in the present
work, we wish to extend our previous analysis~\cite{Gioutsos:2005uw} ( including
also the models proposed in~\cite{Antoniadis:2004dt}), and work out various
interesting phenomenological issues which reveal the fundamental differences for
their non-supersymmetric, supersymmetric and split supersymmetry versions.   We
thus start with a classification of the various D-brane derived models with
Standard Model gauge symmetry extended by $U(1)$ factors. In particular, we
analyse various interesting phenomenological issues of the brane configurations
with $U(3)\times U(2)\times U(1)^N$ symmetry, where the number of additional
$U(1)$ branes is at most three, $N=1,2,3$, so that an economical Higgs sector
arises. We calculate the string unification scale, we discuss the successful
$b-\tau$ equality of the traditional Grand Unified Theories (GUTs) and calculate
the gaugino masses for a wide split-susy range and different initial conditions.
Moreover, we seek preferable solutions that accommodate a right-handed neutrino
$\nu^c$ with appropriate mass (which is usually of the order of the string
scale), so that the `see-saw' mechanism results to a sub-eV effective Majorana
mass scale for the left-handed neutrino component $\nu_L$, as required by
experimental
 and cosmological data. We further examine how these models are discriminated by their
different  predictions for the gaugino masses and the life-time of the gluino.

The paper is organized as follows: In section 2 we present the general set up of
the D-brane constructions based on the $U(3)\times U(2)\times U(1)^N$ symmetry.
We consider all possible configurations with $N=1,2,3$ abelian branes and find
the different hypercharge embeddings compatible with the SM  particle spectrum.
In section 3 we perform a one-loop RG analysis to calculate the string scale for
the models obtained in section 2, both for the case of parallel as well as
intersecting brane scenarios. In section 4 we discuss the fermion mass relations
of the third generation and in particular we examine the possibility of obtaining
the correct bottom-tau mass ratio for $b-\tau$ Yukawa equality at the string
scale. Gaugino masses and the lifetime of the gluino are discussed in section 5,
while in section 6 we draw our conclusions.

\section{Standard Model-like D-brane configurations with extra abelian factors}

We consider models with SM gauge symmetry extended by several $U(1)$ factors
which arise in the context of certain D-brane configurations. The basic
ingredient of such D-brane constructions is the brane stack, i.e.~a certain
number of parallel D-branes which sit at the same point in the transverse space.
A single D-brane carries a $U(1)$ gauge symmetry, while a stack of $n$ parallel
branes gives rise to a $U(n)$ gauge theory where its gauge bosons correspond to
open strings having both their ends attached to the various brane stacks. In flat
space $D$-branes lead to non-chiral matter whilst chirality arises when they are
wrapped on a torus~\cite{Blumenhagen:2000wh,Aldazabal:2000cn}. Furthermore, in
the case of intersecting branes chiral fermions sit in singular points in the
transverse space while the number of fermion generations, and other fermions, are
related to the two distinct numbers of brane wrappings around the two circles of
the torus. We note in passing that the intersecting branes are the T-dual picture
of D-brane case where one turns on magnetic fields~\cite{Bachas:1995ik} to
stabilize the closed string moduli. For two stacks $n_a, n_b$, the gauge group is
$U(n_a)\times U(n_b)$ while the fermions (which live in the intersections) belong
to the bi-fundamental representations $(n_a,\bar n_b)$, or $(\bar n_a, n_b)$.

In our particular constructions the non abelian part of the gauge group is chosen
to be the minimal one needed to accommodate the $SU(3)_C$ and $SU(2)_L$ gauge
 symmetries of the SM. We further assume the existence of $N$ extra $U(1)$
 abelian branes, thus the full gauge group is
\ba%
G={U(3)}_C\times{U(2)}_L\times{U(1)}^N \label{ggg}
\ea%
Since $U(n)\sim {SU(n)}\times{U(1)}$, the particular D-brane construction
automatically gives rise to models with $SU(3)\times SU(2)\times U(1)^{N+2}$
gauge group structure, while SM fermions may carry additional quantum numbers
under these extra $U(1)$'s. Thus, a general observation on the derivation of SM
from D-branes, even in its simplest realization, is that, besides the linear
combination related to the standard hypercharge factor, several $U(1)$ factors
are involved. Many of these $U(1)$'s are anomalous, however, their anomalies are
canceled by a generalized Green-Schwartz mechanism. The corresponding gauge
bosons eventually become massive and the associated $U(1)$ gauge symmetries are
broken. Some of these $U(1)$'s are associated with the conservation of fermion
numbers. It can be seen that, when quarks and leptons belong to the
bifundamentals,  the abelian factor obtained from $U(3)_C\ra SU(3)_C\times
U(1)_C$ is related to the baryon number since all quarks, which transform
non-trivially under the color gauge group, have the same `charge' under $U(1)_C$.
Even when the corresponding gauge symmetry is broken, a global symmetry remains
at low energies which eliminates various baryon number violating
operators~\cite{Antoniadis:2002en,Coriano:2005js}.

The existence of  many $U(1)$ factors allows in principle various embeddings of
the hypercharge, provided that the SM spectrum arises with the correct `charges'
under these embeddings. For the case considered in the present work, the most
general hypercharge gauge coupling condition can be written as follows\footnote{
We have used the traditional normalization ${\rm Tr}\,T^a\,T^b=\delta^{ab}/2,
a,b=1,\dots,n^2$ for the $U(n)$ generators and assumed that the vector
representation (${\bf n}$) has abelian charge $+1$ and thus the $U(1)$ coupling
becomes ${g_n}/{\sqrt{2 n}}$ where $g_n$ the $SU(n)$ coupling.}%
\ba%
\frac{1}{g_Y^2}=\frac{6 k_3^2}{g_3^2}+\frac{4 k_2^2}{g_2^2}+2\sum_{i=1}^N
\frac{{k'_i}^2}{{g'_i}^2}\label{gY}%
\ea%
For a given hypercharge embedding the $k_i'$'s can be determined and equation
(\ref{gY}) relates the weak angle $\sin^2\theta_W=({1+k_Y})^{-1}$ with the gauge
coupling ratios at the string scale $(M_S)$ by:%
\ba
k_Y&\equiv&\frac{\alpha_2}{\alpha_Y}
   \;=\;{6 k_3^2}\,\frac{\alpha_2}{\alpha_3}+4 k_2^2+2\sum_{i=1}^N
{k_i'}^2\,\frac{\alpha_2}{\alpha_i'}\label{kY}
\ea%
where $\alpha_i \equiv g_i^2/(4\pi)$. Note that, in a D-brane context the gauge
couplings do not necessarily attain a common value at the string (brane) scale,
so in general, the ratios $\alpha_2/\alpha_3$, $\alpha_2/\alpha_i'$ differ from
unity there. However, in some classes of models several relations enter between
the gauge couplings at $M_S$ and a partial unification is feasible.

We start our investigation considering the case where some $U(1)$ branes are
superposed with the $U(3)$ stack, whilst the remaining ones are aligned with the
$U(2)$ stack. Therefore, if we assume that $r$ $(r<N)$ parallel $U(1)$ branes
(with $\alpha'_{1,...,r}$ couplings)  are aligned with the $U(3)$ brane, while
the remaining $N-r$ $U(1)$ branes (with $\alpha'_{r+1,...,N}$ couplings) are
superposed with the $U(2)$ brane, this implies at $M_S$ that the couplings
$\alpha'_{1,...,r}$ are equal to $\alpha_{3}$, while $\alpha'_{r+1,...,N}$ are
equal to $\alpha_2$. We then can write $k_Y$ at the string scale as%
\ba%
k_Y&\equiv&\frac{\alpha_2(M_S)}{\alpha_Y(M_S)}\;=\;n_1\,\xi\,+n_2,\label{kY1}
\ea%
where we have defined the gauge coupling ratio $\xi$ entering in (\ref{kY1}) by%
\ba%
\xi&=&\frac{\alpha_2(M_S)}{\alpha_3(M_S)}
\ea%
and the coefficients $n_1,n_2$ by%
\ba%
n_1={6 k_3^2}+2\sum_{i=1}^n {k_i'}^2, ~~~~~~~~~ n_2=4 k_2^2+2\sum_{i=n+1}^N
{k_i'}^2.\label{n1n2}
\ea%

In our D-brane construction each SM particle  corresponds to an open string
stretched between pairs of $U(3)$, $U(2)$ and extra $U(1)'_i$ brane sets and
belongs to some bi-fundamental representation of the associated unitary groups.
Yet, higher antisymmetric or symmetric representations could also be obtained by
considering strings with both ends on the  same brane set. For example, when only
one abelian brane is included beyond the non-abelian ones, some of the Standard
Model fermions should be accommodated in these antisymmetric
representations~\cite{Antoniadis:2004dt}. On the contrary for the cases $N=2,3$
we will see that all known fermions can  be accommodated solely to bi-fundamental
representations. More abelian branes could also be added, however, we restrict
here only to $N\le 3$ in order to ensure an economical Higgs sector and implement
a $b-{\tau}$ Yukawa unification at $M_S$, where down quarks and leptons acquire
their masses from a common Higgs.

\subsection{SM D-Brane configurations with one $ U(1)$ brane}

\begin{figure}[!t]
\centering
\includegraphics[width=0.32\textwidth]{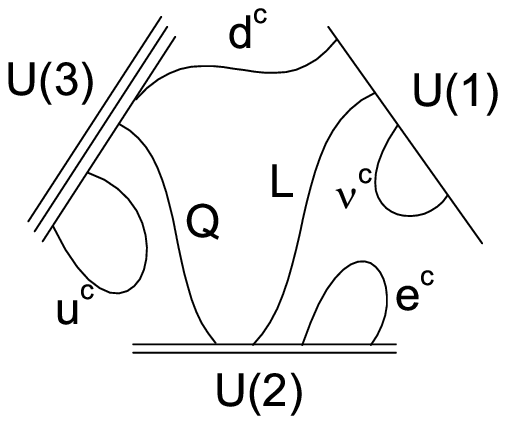}
\includegraphics[width=0.32\textwidth]{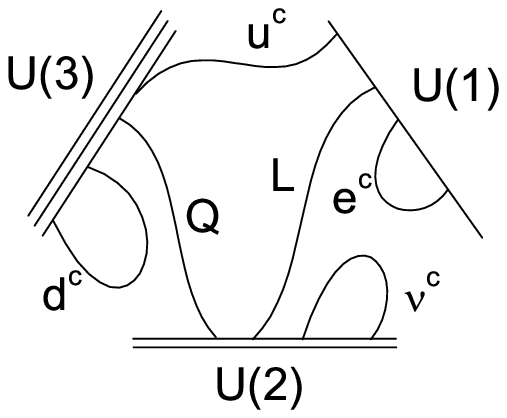}
\includegraphics[width=0.32\textwidth]{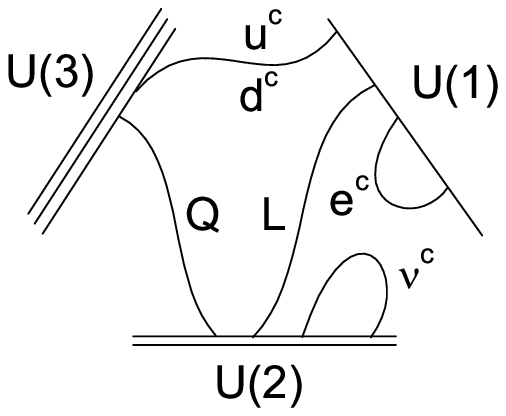}
\caption{\label{f1}Standard Model D-brane configurations with  one
abelian brane ($N=1$).}
\end{figure}

We start our analysis with the minimal gauge symmetry obtained when only one
abelian brane  ($N=1$) is included beyond the $U(3),U(2)$ stacks. Indeed, in
\cite{Antoniadis:2004dt,Rizos} it was shown that the embedding of the SM can be
realized in a minimal set of only three brane-stacks. Three distinct models
$\alpha_1, b_1, c_1$ were constructed in the context of $U(3)\times U(2)\times
U(1)$ symmetry, represented by the three brane configurations shown in
Fig.~\ref{f1}.
In all three cases, quark and lepton doublets are represented by open strings
with ends attached appropriately on the corresponding brane-stacks. In model
$a_1$ the $u^c$ arises from a string with both ends attached to the $U(3)$
brane-stack, thus it belongs to the antisymmetric representation, while $d^c$ is
represented by a string stretched between the color and the $U(1)$ brane. In
model $b_1$ the roles of $u^c$ and $d^c$ are exchanged, while in the last model
$c_1$, both $u^c$ and $d^c$ originate from strings stretched between $U(3)$ and
$U(2)$ stacks. Moreover, the right handed electron is obtained from strings with
both ends attached on $U(2)$ or the $U(1)$ brane-stack as shown in the same
Fig.~\ref{f1}. The quantum number of the SM fermion fields, for these
configurations, are expressed in terms of the sign ambiguities $\eps_i$ ($\eps_i
= \pm 1$, $i=1, \cdots, 6$) and presented in Table \ref{tn1}.
\begin{table}[!b]
\centering
\begin{tabular}{l|ccccc}
&$Q$&$d^c$&$u^c$&$L$&$e^c$\\
\hline\\[-0.38cm]
 $a_1$:&$(3,2;1,\eps_1,0)$&$(\bar 3,1;-1,0,\eps_2)$&$(\bar
3,1;2\eps_3,0,0)$&$(1,2;0,\eps_4,\eps_5)$&$(1,1;0,2\eps_6,0)$
\\
$b_1$:&$(3,2;1,\eps_1,0)$&$(\bar 3,1;2\eps_2,0,0)$&$(\bar
3,1;-1,0,\eps_3)$&$(1,2;0,\eps_4,\eps_5)$&$(1,1;0,0,2\eps_6)$
\\
$c_1$:&$(3,2;1,\eps_1,0)$&$(\bar 3,1;-1,0,\eps_2)$&$(\bar
3,1;-1,0,\eps_3)$&$(1,2;0,\eps_4,\eps_5)$&$(1,1;0,0,2\eps_6)$
\end{tabular}
\caption{The quantum numbers of the SM fermions in the three possible cases which
arise in the $U(3)\times U(2)\times U(1)$ brane configuration.\label{tn1} }
\end{table}
We stress here that the addition of the right handed neutrino ($\nu^c$) forces
some of the  $\eps_i$ to take specific signs, however in what follows we solve
the more general system without the right handed neutrino.

The hypercharge assignment conditions for these configurations determine the
coefficients $k_3,k_2,k_1'$ which subsequently determine  $k_Y$ given by equation
(\ref{kY}). In particular, for the model $a_1$, these are
\[%
\begin{array}{c}%
\displaystyle%
k_3+\eps_1 k_2 = \frac 16 \qquad -k_3+\eps_2 k'_1= \frac 13\\
\displaystyle%
k_2 \eps_4 + k'_1 \eps_5 = -\frac 12 \qquad 2k_3
\eps_3=-\frac 23 \qquad 2\eps_6 k_2=1
\end{array}%
\]%
Solving the above linear system of equations, we find $ k_3= \frac 13, k_2=\frac
12,k_1'=0 $, thus $ k_Y=\frac{2}{3}\xi+1$ which further implies
$\sin^2\theta_W(M_S)=\frac{3}{6+2\xi}$.

Similarly, for the models $b_1$, $c_1$, solving the corresponding linear system
of hypercharge equations one obtains identical solutions $ k_3=\frac 16,
k_2=0,k_1'=\frac 12$ while, $k_Y = \frac 16 \xi + \frac 12
\frac{\alpha_2}{\alpha_1'} $. In the parallel brane scenario, depending on the
orientation of the $U(1)$ brane, $k_Y$ may obtain two distinct values. Thus, if
$\alpha_1'=\alpha_2$ at $M_S$, then $k_Y=\frac{\xi}6+\frac 12$, whilst, if
$\alpha_1'=\alpha_3$, then $k_Y=\frac{2\xi}3$. The above results, are summarized
in the first two rows of Table \ref{t1}.

\subsection{D-Brane configurations with $ N=2,3 $ abelian branes}

\begin{figure}[!t]
\centering
\includegraphics[width=0.3\textwidth]{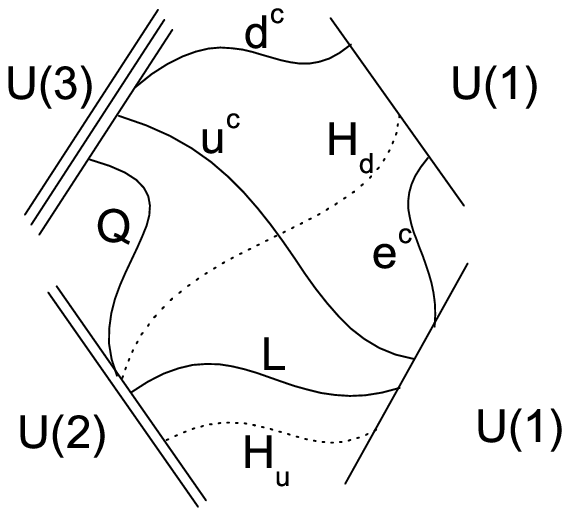}
\includegraphics[width=0.3\textwidth]{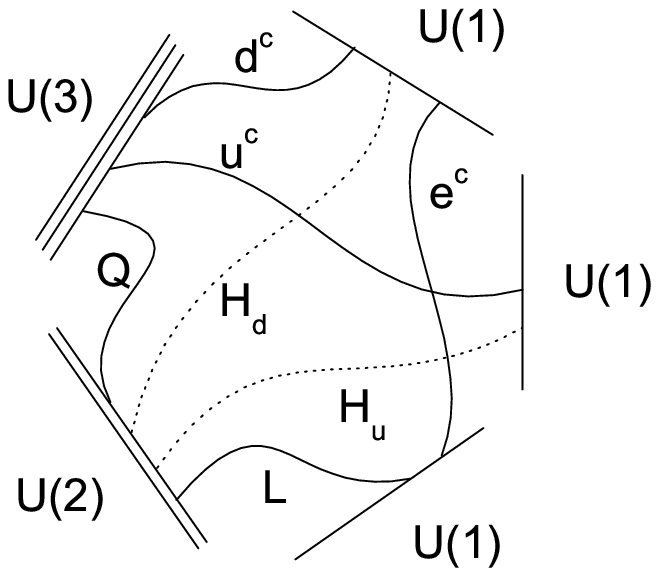}
\includegraphics[width=0.3\textwidth]{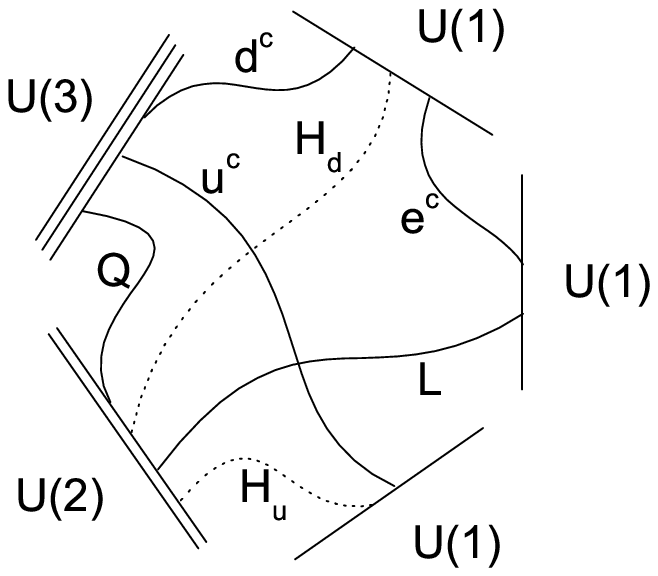}
\caption{\label{fconf}Configurations with two and three extra abelian branes
$N=2,3$.}
\end{figure}

We next study two more brane-configurations, namely those with $N=2$ and $3$
abelian branes in addition to the $U(3), \ U(2)$ stacks, since as explained
above, these are the only ones which share the property for a natural $b-\tau$
unification through an economical Higgs sector. The corresponding configurations
are shown in Figure \ref{fconf}. As stated previously, the possible hypercharge
embeddings for each configuration can be obtained by solving the hypercharge
assignment conditions for SM particles. The SM particle quantum numbers under the
full gauge group $SU(3)\times SU(2)\times U_3(1) \times U_2(1) \times {U(1)'}_1
\times {U(1)'}_2$ are $Q(3,2;+1,\varepsilon_1,0,0)$, $d^c(\bar
3,1;-1,0,\varepsilon_2,0)$, $u^c(\bar 3,1;-1,0,0,\varepsilon_3)$,
$L(1,2;0,\varepsilon_4,0,\varepsilon_5)$,
$e^c(1,1;0,0,\varepsilon_6,\varepsilon_7)$. A similar assignment can be written
also for $N=3$. Then, the hypercharge assignment equations for both brane
configurations can be written
\begin{equation}%
\begin{array}{c}%
\displaystyle%
k_3+k_2\, \e_1 =  \frac 16 \qquad -k_3+k_1'\, \e_2  = \frac 13\\
\displaystyle%
-k_3+k\, \e_3  =  -\frac 23 \qquad k_2 \, \e_4 + k_2'\, \e_5  =  -\frac 12 \qquad
k_1' \, \e_6 + k_2'\, \e_7  = 1
\label{6e}
\end{array}%
\end{equation}%
where as previously $\e_i=\pm1$, $i=1,\cdots,7$ . Here we have used a compact
notation where $k=k_2'$ for the $N=2$ configuration and $k=k_3'$ for the $N=3$
one. For the $N=2$ case, the system (\ref{6e}) is satisfied by two sets
of solutions%
\ba
 |k_3|=|\frac 23 -k_2'| \ \ , \ \ |k_2|=|\frac 12 -k_2'| \ \ , \ \
 |k_1'|=|1-k_2'| \ \ , \ \ k_2' \label{discrn=01}\\
 |k_3|=|\frac 23 +k_2'| \ \ , \ \ |k_2|=|\frac 12 +k_2'| \ \ , \ \
 |k_1'|=|1+k_2'| \ \ , \ \ k_2'\label{discrn=02}
\ea%
while for the $N=3$ case, there are also two  sets of solutions as well as one
`discrete' solution which cannot be obtained from the other two%
\ba%
 |k_3|=|\frac 23 -k_3'| \ \ , \ \ |k_2|=|\frac 12 -k_3'| \ \ , \ \
 |k_1'|=|1-k_3'| \ \ , \ \ |k_2'|=|k_3'| \ \ , \ \ k_3' \label{discrn=1}\\
 |k_3|=|\frac 23 +k_3'| \ \ , \ \ |k_2|=|\frac 12 +k_3'| \ \ , \ \
 |k_1'|=|1+k_3'| \ \ , \ \ |k_2'|=|k_3'| \ \ , \ \ k_3' \label{discrn=2}\\
 |k_3|=\frac 56 \ \ , \ \ |k_2|=1 \ \ , \ \
 |k_1'|= \frac 12 \ \ , \ \ |k_2'|=\frac 12 \ \ , \ \ k_3'=\frac 32.
 \label{discrn=3}
\ea%
\begin{table}[!t]
\centering
\renewcommand{\arraystretch}{1.2}
\begin{tabular}{|c|l|c|c|c|c|c|}
\hline
$N$ & & $|k_3|$ & $|k_2|$ & $|k_1'|$ & $|k_2'|$ & $|k_3'|$\\
\hline
    & $a_1$ & $\frac 13$ & $\frac 12$ & $0$        & $-$ & $-$\\
\raisebox{1.5ex}[0pt]{$1$} & $b_1$ & $\frac 16$ & $0$        & $\frac 12$ & $-$ & $-$\\
\hline
    & $a_2$ & $\frac 16$ & $0$        & $\frac 12$ & $\frac 12$ & $-$\\
$2$ & $b_2$ & $\frac 23$ & $\frac 12$ & $1$        & $0$        & $-$\\
    & $c_2$ & $\frac 13$ & $\frac 12$ & $0$        & $1$        & $-$\\
\hline
    & $a_3$ & $\frac 16$ & $0$        & $\frac 12$ & $\frac 12$  & $\frac 12$\\
$3$ & $b_3$ & $\frac 13$ & $\frac 12$ & $0$        & $1$         & $1$\\
    & $c_3$ & $\frac 23$ & $\frac 12$ & $1$        & $0$         & $0$\\
\hline
\end{tabular}
\caption{\label{t1}The simplest hypercharge embeddings for SM
D-brane configurations with  $N=1,2,3$ abelian brane.}
\end{table}
As can be seen, solutions (\ref{discrn=01})-(\ref{discrn=2}) are expressed in
terms of one free parameter, namely $k_2'$ or $k_3'$ for $N=2$, $3$ respectively.
Further, it can be checked that solution (\ref{discrn=3}), leads to a physically
unacceptable model and it will not be elaborated further.

To start our analysis, we pick up specific values for $k_2', k_3'$ which imply
the simplest solutions of equations (\ref{discrn=01})-(\ref{discrn=2}). Hence,
choosing $k_2'=  \frac 12, 0, 1$ and $k_3'=  \frac 12, 1, 0$, we end up with the
models $a_2$, $b_2$, $c_2$ and $a_3$, $b_3$, $c_3$ presented in Table \ref{t1}.
Note that trivial values of $k, k'$ indicate that the associated abelian factor
does not contribute to the hypercharge. In the following sections we will analyze
the predictions of these models for the string scale, the bottom-tau unification
and the gaugino masses. We will further extend our analysis to more general gauge
coupling relations at $M_S$ which may occur in intersecting brane scenarios.

\section{Unification and the String Scale\label{unif}}

One of the most interesting properties of the SM gauge couplings, which led to
the exploration  of supersymmetric Grand Unified Theories (GUTs), is the fact
that when they  are extrapolated at high energies, they merge to a common
coupling at a scale of the order $10^{16}$ GeV. In addition, traditional GUTs
imply  a value for the weak mixing angle, which gives low energy predictions in
agreement  with the experiment. In D-brane constructions however, the  SM gauge
couplings do not necessarily satisfy the usual unification condition. The reason
is that in this case, the volume of the internal space enters between gauge and
string couplings, thus the actual values of the SM gauge couplings may differ at
the string scale. Nonetheless, in certain classes of D-brane configurations it is
possible that some internal volume relations allow for a partial unification.
Such configurations at the string level arise from the superposition of the
associated parallel brane stacks. Other particular gauge coupling relations may
also arise in classes of intersecting brane models.
\newc{\rsb}[2]{\raisebox{#1}[0pt]{#2}}
\def\m3#1#2#3{($\mathbf{#1}, \mathbf{#2}, \mathbf{#3}$)}
\begin{table}[!t]
\centering
\renewcommand{\arraystretch}{1.3}
\begin{tabular}{|c|c|c|c|}
\hline%
  & $a_1$ & \multicolumn{2}{|c|}{$b_1$}\\
\hline%
   & $\bf 2$ &         &          \\
\rsb{1.5ex}{$U'(1)_1$} & $\bf 3$ & \rsb{1.5ex}{$\bf 2$} & \rsb{1.5ex}{$\bf 3$}  \\
\hline%
($n_1$, $n_2$)& $(\frac 23, 1)$ & $(\frac 16, \frac 12)$ & $(\frac 23, 0)$ \\
\hline%
\multicolumn{4}{c}{}
\end{tabular}
\begin{tabular}{|c|c|c|c|c|c|c|c|}
\hline%
  & \multicolumn{3}{|c|}{$a_2$} & \multicolumn{2}{|c|}{$b_2$} & \multicolumn{2}{|c|}{$c_2$}\\
\hline%
 &   &  &
$(\mathbf{2},\mathbf{3})$ & $(\mathbf{2},\mathbf{2})$ & $(\mathbf{3},\mathbf{3})$
& $(\mathbf{2},\mathbf{2})$ & $(\mathbf{2},\mathbf{3})$ \\
\rsb{1.5ex}{$U'(1)_{\{1,2\}}$} & \raisebox{1.5ex}[0pt]{$(\mathbf{2},\mathbf{2})$} &
\raisebox{1.5ex}[0pt]{$(\mathbf{3},\mathbf{3})$} & $(\mathbf{3},\mathbf{2})$ &
$(\mathbf{2},\mathbf{3})$ & $(\mathbf{3},\mathbf{2})$
& $(\mathbf{3},\mathbf{2})$ & $(\mathbf{3},\mathbf{3})$ \\
\hline%
($n_1$, $n_2$)& $(\frac 16, 1)$& $(\frac 76, 0)$&
      $(\frac 23, \frac 12)$&
      $(\frac 83, 3)$&
      $(\frac{14}{3}, 1)$& $(\frac 23, 3)$ & $(\frac 83, 1)$\\
\hline%
\multicolumn{8}{c}{}
\end{tabular}
\setlength{\tabcolsep}{0.8mm}
\begin{tabular}{|c|c|c|c|c|c|c|c|c|c|}
\hline%
  & \multicolumn{4}{|c|}{$a_3$} & \multicolumn{3}{|c|}{$b_3$} & \multicolumn{2}{|c|}{$c_3$}\\
\hline%
 & & & & & & & \m3 223 & \m3 222 & \m3 333 \\
 & & \rsb{1.5ex}{\m3 232} & & \rsb{1.5ex}{\m3 323} & \m3 222 & \m3 233
     & \m3 232 & \m3 223 & \m3 322 \\
 \rsb{1.5ex}{$U'(1)_{\{1,2,3\}}$} & \rsb{1.5ex}{\m3 222} & \rsb{1.5ex}{\m3 322} & \rsb{1.5ex}{\m3 333}
     & \rsb{1.5ex}{\m3 332} & \m3 322 & \m3 333 & \m3 323 & \m3 232 & \m3 323 \\
 & & \rsb{1.5ex}{\m3 223} & & \rsb{1.5ex}{\m3 233} & & & \m3 332 & \m3 233 & \m3 332 \\
\hline%
($n_1$, $n_2$)& $(\frac 16, \frac 32)$& $(\frac 23, 1)$
      & $(\frac 53, 0)$ & $(\frac 76, \frac 12)$
      & $(\frac 23, 5)$& $(\frac{14}{3}, 1)$ & $(\frac 83, 3)$
      & $(\frac 83, 3)$ & $(\frac{14}{3}, 1)$\\
\hline%
\end{tabular}
\caption{\label{ytab2A} Various $U(1)'$ alignments and corresponding $n_{1,2}$
values which determine $k_Y=n_1\xi+n_2$ for all models in Table \ref{t1}. The
first part of the Table refers to the case where only one $U(1)$ brane is
included in the configuration. Similarly, the second and third parts of the Table
correspond to configurations with two and three additional $U(1)$ branes.}
\end{table}

Particularly, for D-brane models with split supersymmetry in the presence of
internal magnetic fields ${\cal H}$, certain requirements for unification of
$\alpha_2$ and $\alpha_3$ gauge couplings have already been discussed in
ref.~\cite{Antoniadis:2004dt}. More precisely, in this case the non-abelian
four-dimensional $\alpha_2$ and $\alpha_3$ gauge couplings are given by%
\ba%
\alpha_i^{-1} &=& \frac{V^i}{g_s} \ \prod_{I=1}^3 |n_I^i|\,
 \sqrt{1+({\cal H}_I^i\alpha')^2}  \label{a23}
\ea%
where $g_s$ is the string coupling, $V^i$ is the compactification volume of the
i-th stack, $n_I^i$ is the number of wrappings around the I-th torus and ${\cal
H}_I^i$ is the corresponding magnetic field component. Equality of $\alpha_2$ and
$\alpha_3$ gauge couplings may occur when: (i) the compactification volume is
independent of the particular brane stack, (ii) there is no multiple wrapping
($|n_I^i|=1$) and (iii) magnetic fields $({\cal H}_I^i\alpha' \ll 1)$ are
sufficiently weak.

In order to examine the more general case, when the magnetic fields are not weak,
we relax the $\alpha_3 = \alpha_2$ condition and follow a different approach. At
the string scale we assume that some $U(1)$ gauge couplings are equal to the
$U(3)$ one, while the remaining $U(1)$ couplings are equal to the $U(2)$ one. A
more general analysis, where the extra $U(1)$ gauge couplings $\alpha_i'$ take
arbitrary values (corresponding to possible  cases of intersecting branes),
follows in the end of this section. We start our analysis by deriving the
one-loop analytic expressions for the string scale in the context of split
supersymmetry, considering various cases of partial gauge coupling unification at
$M_S$. We express $M_S$ in terms of the low energy experimentally measured values
$\alpha_3,\alpha_e$, $\sin^2\theta_W$ and determine its range as a function of
the split-supersymmetry scale $\tilde m$ (an average scale for the scalar
supersymmetric spectrum) by imposing eq.~(\ref{kY}). We discuss how the string
scale predictions discriminate models with different numbers of $U(1)$ branes, as
well as different hypercharge embeddings and gauge coupling relations at $M_S$.
For the sake of completeness, we also compare the results with the supersymmetric
and non-supersymmetric cases.

Following closely the analysis in ref.~\cite{Gioutsos:2005uw}, we first
concentrate on all possible gauge coupling relations arising from various models
in the context of non-intersecting branes. Consequently, we are led to a discrete
number of admissible cases which are presented in Table \ref{ytab2A}. Thus, when
only one $U(1)$ brane is included in the configuration, this can be superposed
either with the $U(2)$ or with the $U(3)$ brane-stack. In model $a_1$, since
$k_1'=0$, both cases lead to the same $k_Y$ value. In model $b_2$, we have
$k_1'=1/2$, hence we obtain two distinct cases presented in the upper part of
Table \ref{ytab2A}. A similar analysis results to the cases presented in the same
Table when two and three abelian branes are included in the configuration. For
clarity we stress here that the notation \m3 223 indicates the orientation of the
extra $U(1)$ branes where $\bf 2$ stands for the $U(2)$-direction and $\bf 3$
stands for the $U(3)$ one. Hence \m3 223 means that the first two abelian branes
in the three abelian brane scenario are aligned with the $U(2)$ stack, while the
third is aligned with the $U(3)$ stack. We proceed now to analyze the cases of
Table \ref{ytab2A}.

\subsection{Correlation of the string and the split SUSY scale}

In order to investigate qualitatively the influence of the split SUSY scale
($\tilde{m}$) on  the string scale magnitude in various models of Table~\ref{t1},
it is sufficient to work out analytic expressions of the gauge coupling RGEs at
the one-loop order (see Appendix). Above $\tilde m$, the beta function
coefficients are those of the full supersymmetric theory, whereas below $\tilde
m$ beta functions have contributions only from the fermion partners (SM fermions,
gauginos and higgsinos) and one linear combination of the scalar Higgs doublets.
We denote by $b_i^{SU}$, $b_i$ the beta-coefficients valid in the energy range
above and below $\tilde m$ respectively. Then, the one-loop RGEs lead to the
following expression for the
string scale%
\ba%
M_S&=& M_Z  \, \left(\frac{\tilde m}{M_Z}\right)^{\frac{\cal N}{\cal
D}}e^{f(n_1,n_2)}\label{stringscale}
\ea%
where the numerator ${\cal N}$ and the denominator ${\cal D}$ of the exponent are%
\ba%
{\cal N}&=& n_1 (b_3-b_3^{SU}) + n_2 (b_2-b_2^{SU}) - (b_Y-b_Y^{SU}) \label{ncap}\\
{\cal D}&=& - n_1 b_3^{SU} - n_2 b_2^{SU} + b_Y^{SU} \label{dcap}
\ea%
On the other hand the function $f(n_1,n_2)$ in (\ref{stringscale}) depends on the
coefficients $n_1, n_2$ and the experimentally measured values for $\alpha_Y$,
$\alpha_2$ and $\alpha_3$ at $M_Z$. In particular\footnote{%
Recall that $\alpha_Y=\frac{\alpha_e}{1-\sin^2\theta_W}$ and
$\alpha_2=\frac{\alpha_e}{\sin^2\theta_W}.$},%
\ba%
f(n_1,n_2) &=& \frac{2\pi}{\cal D} \left( \frac{1}{\alpha_Y} -
\frac{n_2}{\alpha_2}- \frac{n_1}{\alpha_3} \right)
\ea%
Substituting the values of $b_i,b_i^{SU}$ (see Appendix) in (\ref{ncap}),
(\ref{dcap}) one finds that the ratio $\mathcal{N}/\mathcal{D}$ is given by%
\ba%
 \frac{\cal N}{\cal D} = \frac{21-12n_1-13n_2}{6(11+3n_1-n_2)}
\ea%

Now, for all models of Table \ref{t1}, the specific values of $n_{1,2}$ can be
calculated  from the coefficients $k_i, k_i'$, using relations (\ref{n1n2}) (see
Table~\ref{ytab2A}). As can be seen from (\ref{stringscale}), for any model of
Table \ref{t1}, the string scale $M_S$ could either increase or decrease as the
split SUSY scale increases. The correlation between $M_S$ and $\tilde m$
certainly depends on the sign of the exponent of the ratio $\tilde m/M_Z$ the
latter being always greater that unity. For the models under consideration, the
denominator ${\cal D}$ turns out to be always positive, therefore, the sign of
the exponent  depends on the sign of the numerator ${\cal N}$ which can be
checked by substituting the values of $n_1,n_2$ obtained for the various models
presented in Table~\ref{ytab2A}.
\begin{figure}[!t] \centering
\includegraphics[width=0.75\textwidth]{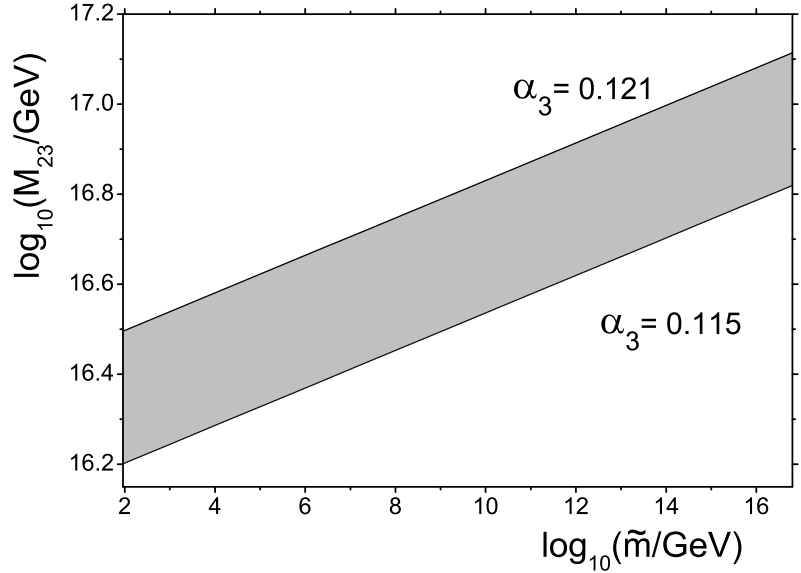}
\caption{The one-loop unification range of $\alpha_{2,3}$ gauge couplings
 as a function of the split SUSY scale. The $\tilde m$-dependence appears
 due to the different number of Higgs doublets above and below $\tilde m$.}
\label{m23.eps}
\end{figure}
However, before we proceed further to classify
the various models with respect to the string scale predictions, we discuss two
naturalness  conditions that should be satisfied. First of all, it is obvious
that the relation (\ref{stringscale}) for the string scale is valid for $\tilde m
\le M_S$. On the other hand, as has already been noted in section \ref{unif}, the
non-abelian gauge couplings do not necessarily unify at the string scale $M_S$,
hence for a given model naturalness imposes the condition $\alpha_2(\mu)\le
\alpha_3(\mu)$ for any scale $\mu\le M_S$. If we denote by
$M_{23}(\tilde{m},\alpha_3)$ the scale where the $\alpha_2(\mu)$, $\alpha_3(\mu)$
gauge couplings merge to a common value, then for the models
under consideration, the following condition should be imposed\footnote{%
At this scale $\alpha_Y$ is smaller than the common $\alpha_{2,3}$ value for
every $\tilde{m}$ which means that always $\alpha_2$, $\alpha_3$ couplings meet
first.}%
\ba%
M_S &\le& M_{23}(\tilde m,\alpha_3)=  M_Z \left( \frac{\tilde{m}}{M_Z}
\right)^{\frac{\mathcal{N}'}{\mathcal{D}'}} e^{f'} \quad\mbox{where}\quad f' =
\frac{2\pi}{\mathcal{D}'} \left( \frac{1}{\alpha_2} - \frac{1}{\alpha_3} \right)
\label{upMS}
\ea%
where ${\cal N}'= (b_2^{SU} - b_3^{SU}) - (b_2 - b_3)$ and  ${\cal
D}'=b_2^{SU}-b_3^{SU}$. In split SUSY with the SM fermion spectrum embedded in
complete $SU(5)$ multiplets, ${\cal N}'$ is proportional to the difference of the
Higgs doublets above and below the split SUSY scale $\tilde m$, thus
 we find $\frac{{\cal N}'}{{\cal D}'}=\frac{\delta n_H}{24}=\frac{1}{24}$.

\begin{table}[!t]
\centering
\renewcommand{\arraystretch}{1.3}
\begin{tabular}{|c|c|c|c|}
\hline%
\multicolumn{2}{|c|}{Case}& Condition & $\tilde{m}$ bound  \\
\hline%
 & $f<f'$ & $\tilde{m} \leq M_Z\, e^{|\omega|}$  & $\leq M_Z\, e^\omega$\\
\cline{2-4}%
\rsb{1.5ex}{$\frac{\cal N}{\cal D} > \frac{1}{24}$}
    & $f>f'$ & $\tilde{m} \leq M_Z\, e^{-|\omega|}$ & $< M_Z$ (Unacceptable) \\
\hline%
 & $f<f'$ & $\tilde{m} \geq M_Z\, e^{-|\omega|}$ &  \\
\cline{2-3}%
\rsb{1.5ex}{$\frac{\cal N}{\cal D} < \frac{1}{24}$} & $f>f'$
    & $\tilde{m} \geq M_Z\, e^{|\omega|}$  & \rsb{1.5ex}{$\geq M_Z\, \max(1,e^{-\omega})$}\\
\hline%
\end{tabular}
\caption{\label{gensys}Bounds imposed on $\tilde{m}$ from naturalness condition
(\ref{natcond}).}
\end{table}
In Figure \ref{m23.eps} we plot $M_{23}$ as a function of $\tilde m$, taking into
account the experimental uncertainties of $\alpha_3$; for given $\tilde m$ we can
check the maximum allowed value of the string scale $M_S$  satisfying the
naturalness criterion $M_S \le  M_{23}$. We further infer from the same figure
that, as $\tilde m$ varies from $M_Z$ to $M_S$,  the $\alpha_2-\alpha_3$
unification point lies in the range (up to two-loop and threshold corrections),
\ba%
 M_{23} &\approx &[1.6 \times 10^{16}-1.2 \times 10^{17}]~
{\rm GeV}.\label{max}%
\ea%
For a more detailed analysis we can use eq.~(\ref{stringscale}) and rewrite
(\ref{upMS}) as a constraint for $\tilde{m}$
\ba%
\left( \frac{\tilde{m}}{M_Z} \right)^{\mbox{sign}\left(\frac{\cal N}{\cal
D}-\frac{1}{24}\right)} &\leq& e^{\omega}\quad\mbox{where}\quad
\omega=-\frac{f-f'}{|\frac{\cal N}{\cal D}-\frac{1}{24}|}. \label{natcond}
\ea%
All possible cases are summarized in Table~\ref{gensys}. Let us further proceed
to present the results obtained for some characteristic cases.

\subsubsection{D-brane Standard Models with one abelian brane.}

We consider first these models of Table \ref{ytab2A} where only one abelian brane
is included in the configuration. There are two different hypercharge embeddings
in the parallel brane scenario (first two rows of Table \ref{fconf}), which
result to three distinct predictions for the string scale (see Table
\ref{ytab3}). Among them, the most interesting one is $a_1$ which predicts a
string scale of the order
\ba%
M_S&=& M_Z \, \exp\left\{\frac{\pi}{6}\left( \frac{1}{\alpha_Y} -
\frac{1}{\alpha_2} - \frac{2}{3\alpha_3} \right) \right\} \approx 2 \times
10^{16}~\mbox{GeV} \label{SGUT}
\ea%
There are two noticeable points that should be mentioned here: firstly, it is a
remarkable fact that in this minimal D-brane configuration, $M_S$ coincides with
the GUT scale obtained in traditional supersymmetric unification
models\footnote{For this configuration the trivial hypercharge normalization $k_Y
= 5/3,\ \xi = 1$ is recovered since from Table \ref{ytab2A} we have $n_1 = 2/3$
and $n_2 = 1$, in accordance with ref.~\cite{Antoniadis:2004dt}.}%
. Secondly, due to the fact that in this case ${\cal N}=0$, the condition
defining the string scale at the one-loop level is satisfied for any $\tilde m$
(a weak dependence is expected at two-loop level). However,  the value of $\tilde
m$ could be fixed if we impose a certain condition on $\alpha_2$, $\alpha_3$ at
$M_S$. In particular, assuming complete unification of the non-abelian gauge
 couplings at the string scale we can use (\ref{upMS}) and determine $\tilde m$
 by the condition $M_{23}(\tilde{m},\alpha_3)\equiv M_S$. It is easy to see that
$$ \tilde{m} = M_Z \exp\left\{ 4\pi \left( \frac{1}{\alpha_Y} -
\frac{4}{\alpha_2} + \frac{7}{3\alpha_3} \right) \right\}$$%
with $ M_Z \leq \tilde{m}  \lesssim 6.26~\mbox{TeV}$ provided that
$\alpha_3 \leq 0.11694$.

The two cases of model $b_1$, for our particular parallel D-brane scenario lead
to higher $M_S$ scales $(M_S \ge M_{Pl})$ therefore, from condition (\ref{max})
and the fact that they lead to a high see-saw suppression of the left handed
neutrino components ($m_{\nu}^{eff}\le 10^{-6}$eV), we infer that none of them
can play the role of a viable low energy effective field theory in the context of
the parallel brane scenario.

\subsubsection{D-brane Standard Models with two abelian branes.}

Three models $a_2,b_2$ and $c_2$ were analyzed for the D-brane configurations
with two extra abelian branes. For the model $a_2$, depending on the orientation
of the $U(1)'$ branes relative to $U(2)$, $U(3)$ stacks, we have three possible
$n_1$, $n_2$ sets (see Table \ref{ytab2A}). In all these cases the string scale
increases as $\tilde m$ attains higher values as well, which can be trivially
inferred from the numerical values of the exponent ${\cal N}/{\cal D}$ namely
$2/21$, $7/87$, $13/150$. In the second case for example, the string scale
\ba%
M_S &=& M_Z \left(\frac{\tilde m}{M_Z}\right)^{\frac{7}{87}} \, \exp\left\{
\frac{4\pi}{29} \left( \frac{1}{\alpha_Y} - \frac{7}{6\alpha_3} \right) \right\}
\label{csa2}
\ea%
turns out to be at least of the order of the Planck mass, even for a split SUSY
scale comparable to the electroweak scale. Similar expressions can be derived for
the other two cases of model $a_2$ (see Table \ref{ytab3}). We should further
note that the naturalness criterion (\ref{max}) discussed in the previous
section, is not satisfied in model $a_2$ since the non-abelian $\alpha_{2,3}$
gauge couplings meet at a scale lower than the one defined by condition
(\ref{kY1}).
\begin{table}[!t]
\centering
\renewcommand{\arraystretch}{1.6}
\begin{tabular}{|c|c|l|c|}
\hline
Models& $(n_1,n_2)$ & \multicolumn{1}{|c|}{$M_S$ (GeV)} & $\tilde{m}$ bound\\
\hline%
$a_1$ &  $(\frac 23,1)$& $[1.90-2.20] \times 10^{16} $ & $\geq M_Z\max(1,\rho)$\\
\hline%
 & $(\frac 23,0)$& $[2.42-2.78] \times 10^{21} \  (\frac{\tilde{m}}{M_Z})^{\frac 16}$ & Unacceptable \\
\rsb{2ex}{$b_1$}&$(\frac 16,\frac 12)$&$[2.12-2.21] \times 10^{22} \ (\frac{\tilde{m}}{M_Z})^{\frac{25}{132}}$ & Unacceptable \\
\hline%
\hline%
 &$(\frac 16,1)$   &$[2.84-2.96] \times 10^{19} \ (\frac{\tilde{m}}{M_Z})^{\frac{2}{21}}$ & Unacceptable \\
$a_2$&$(\frac 76,0)$ & $[3.60-4.48] \times 10^{18} \ (\frac{\tilde{m}}{M_Z})^{\frac{7}{87}}$ & Unacceptable \\
&$(\frac 23,\frac 12)$ &$[8.57-9.90] \times 10^{18} \  (\frac{\tilde{m}}{M_Z})^{\frac{13}{150}}$ & Unacceptable \\
\hline%
 &$(\frac 83,3)$  &$[4.42-6.94] \times 10^{-1} \  (\frac{\tilde{m}}{M_Z})^{-\frac{25}{48}}$ & $\geq M_Z$\\
\rsb{2ex}{$b_2$}&$(\frac{14}{3},1)$ & $[1.46-2.47] \times 10^{5} \ (\frac{\tilde{m}}{M_Z})^{-\frac{1}{3}}$ & $\geq M_Z$\\
\hline%
 &$(\frac 23,3)$  & $[1.00-1.20] \times 10^3 \ (\frac{\tilde{m}}{M_Z})^{-\frac{13}{30}}$ & $\geq M_Z$\\
\rsb{2ex}{$c_2$}&$(\frac{8}{3},1)$  &$[7.39-11.0] \times 10^{8} \ (\frac{\tilde{m}}{M_Z})^{-\frac{2}{9}}$ & $\geq M_Z$\\
\hline%
\hline%
 & $(\frac 16,\frac 32)$  &$ [1.96-2.05] \times 10^{16} \  (\frac{\tilde{m}}{M_Z})^{-\frac{1}{120}}$ & \\
 & $(\frac{2}{3},1)$  & $ [1.90-2.20] \times 10^{16}$ & \\
\rsb{2ex}{$a_3$} & $(\frac 53,0)$ & $ [1.81-2.41] \times 10^{16} \ (\frac{\tilde{m}}{M_Z})^{\frac{1}{96}}$ & \rsb{2ex}{$\geq M_Z\max(1,\rho)$}\\
 & $(\frac 76,\frac 12)$  &$[1.85-2.32] \times 10^{16} \ (\frac{\tilde{m}}{M_Z})^{\frac{1}{168}}$ & \\
\hline
 &$(\frac{2}{3},5)$  &$ [1.23-1.54] \times 10^{-17} \ (\frac{\tilde{m}}{M_Z})^{-\frac{13}{12}}$ & $\geq M_Z$\\
$b_3$ & $(\frac{14}{3},1)$  &$[1.46-2.47] \times 10^{5} \ (\frac{\tilde{m}}{M_Z})^{-\frac{1}{3}}$ & $\geq M_Z$\\
 & $(\frac 83,3)$  & $[4.42-6.94] \times 10^{-1} \ (\frac{\tilde{m}}{M_Z})^{-\frac{25}{48}}$ & $\geq M_Z$\\
\hline
 &$(\frac 83,3)$  &$ [4.42-6.94] \times 10^{-1} \ (\frac{\tilde{m}}{M_Z})^{-\frac{25}{48}}$ & $\geq M_Z$\\
 \rsb{2ex}{$c_3$} &$(\frac{14}{3},1)$  &$[1.46-2.47] \times 10^{5} \ (\frac{\tilde{m}}{M_Z})^{-\frac{1}{3}}$ & $\geq M_Z$\\
 \hline
\end{tabular}
\caption{\label{ytab3} The prediction of the string scale for the various models
of Table \ref{ytab2A} where $\ln\rho = 4\pi \left( \frac{1}{\alpha_Y} -
\frac{4}{\alpha2} + \frac{7}{3\alpha_3} \right)$ and $\rho \in [2.217\times
10^{-4}, 68.636]$ due to $\alpha_3$ uncertainties. Last column shows the $\tilde
m$-bound due to the naturalness condition (\ref{upMS}).}
\end{table}

Likewise, there are two cases in model $b_2$. As can be checked from Table
\ref{ytab3}, the first case leads to an unacceptable small string scale, while
the second one predicts a string scale at the TeV range, given by%
\ba%
M_S &=& M_Z \, \left(\frac{\tilde{m}}{M_Z}\right)^{-\frac{1}{3}} \, \exp\left\{
\frac{\pi}{12} \left( \frac{1}{\alpha_Y} - \frac{1}{\alpha_2} -
\frac{14}{3\alpha_3} \right)\right\}. \label{tev}
\ea%
In (\ref{tev}), the exponent of the mass ratio $\tilde m/M_S$ turns out to be
negative,  thus, as the scale $\tilde m$ decreases, the string scale $M_S$
increases. In the extreme case where $\tilde m\sim M_Z$, i.e., when the split
susy scale is comparable to the electroweak scale,  its highest value is $M_S
\sim 100$ TeV, while for $\tilde m\sim M_S$, the string scale becomes as low as
$M_S\sim 23.5$ TeV.

Two more cases arise in model $c_2$. Again, the first one predicts unacceptably
small string scale. However, the second case of $c_2$ is more interesting. There
$M_S$ is given by the expression
$$M_S \, =\,M_Z \, \left(\frac{\tilde{m}}{M_Z}\right)^{-\frac{2}{9}} \,
\exp\left\{ \frac{\pi}{9} \left( \frac{1}{\alpha_Y} - \frac{1}{\alpha_2}
-\frac{8}{3 \alpha_3} \right)\right\}$$%
which predicts an intermediate string scale of the order $\sim 10^8$ GeV (see
Table \ref{ytab3}). We will see in the next sections that in this model the low
energy $m_{b}/m_{\tau}$ ratio is compatible with a $b-\tau$ Yukawa unification at
$M_S$.

\subsubsection{ D-brane Standard Models with three abelian branes.}

We focus on three classes of models $a_3,b_3,c_3$ which have the following
features. Depending on the various orientations of the extra abelian branes,
model $a_3$ occurs in four realizations and predicts a string scale compared to
the traditional SUSY GUT scale $10^{16}$ GeV. In particular, the string scale for
the second case of $a_3$ is given by eq.~(\ref{SGUT}), leading to an identical
prediction with the simpler model $a_1$. The remaining three cases of $a_3$
exhibit only a weak dependence on the scale $\tilde m$. Finally, models $b_3,c_3$
give similar values for $M_S$ to those of model $b_2$. The results found for all
cases, are summarized  in Table \ref{ytab3}.
\begin{figure}[!t]
\centering
\includegraphics[width=0.49\textwidth]{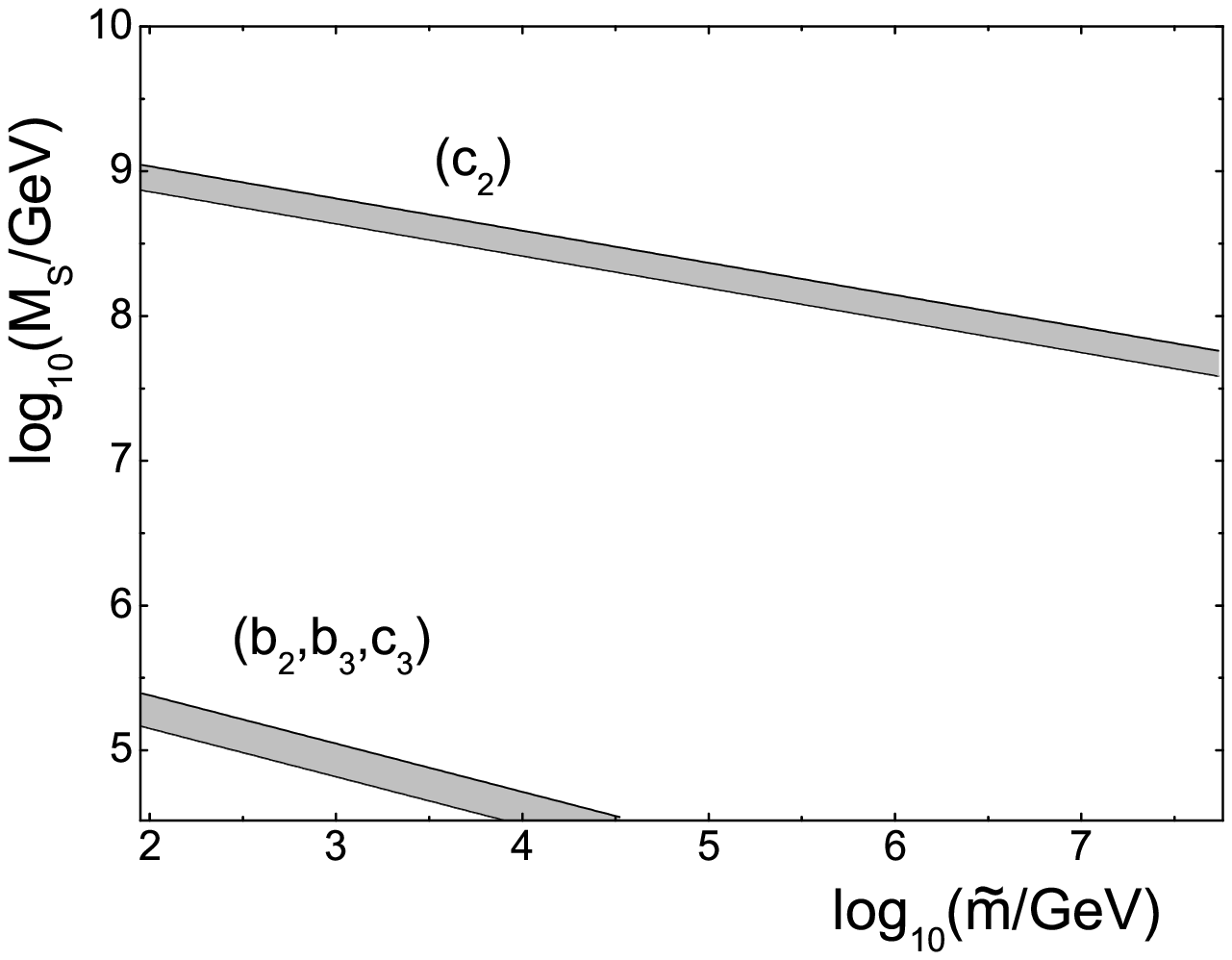}
\includegraphics[width=0.49\textwidth]{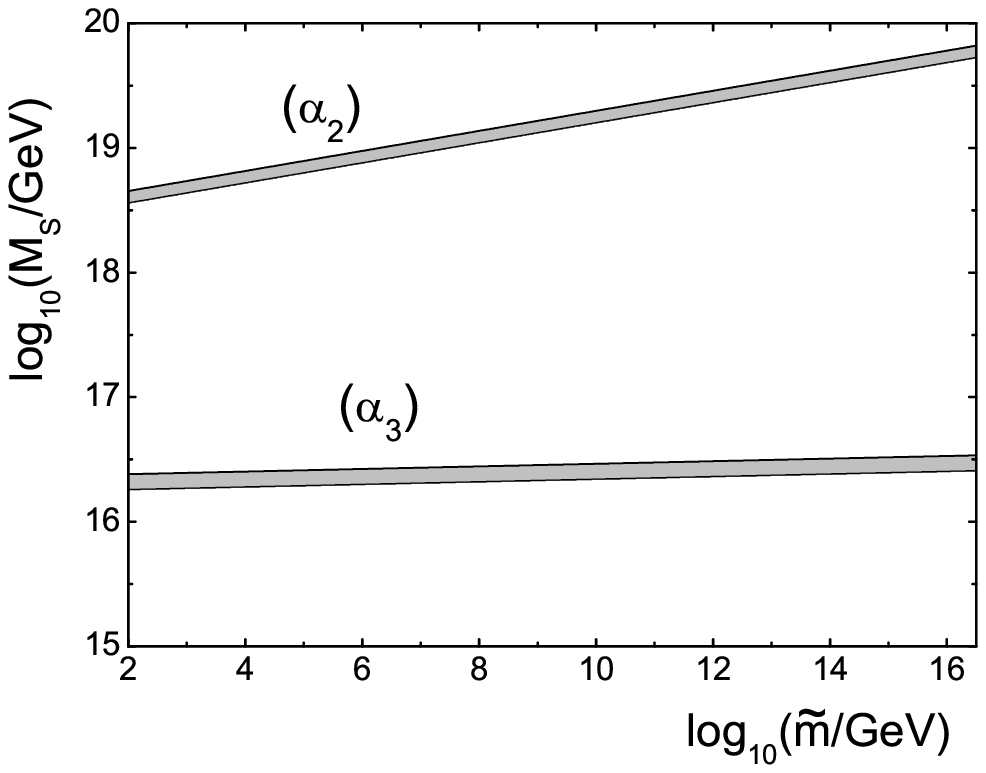}
\caption{\label{ms}The string scale $M_S$ vs the split susy scale $\tilde{m}$ in
the parallel brane scenario for the four characteristic cases discussed in the
text. The thickness of the $M_S$ curves takes into account the experimental
uncertainties of the strong gauge coupling at $M_Z$.}
\end{figure}%

In order to see some qualitative features of $M_S$ we plot in Fig.~\ref{ms} the
string scale $M_S$ versus the split-susy scale $\tilde m$ for four characteristic
cases. We observe from this plot, that for brane configurations with certain
$U(1)$ orientations, which allow for a low (TeV) or an intermediate string scale
(left part of the graph), $M_S$ decreases as $\tilde m$ increases. Models with
low $M_S$ exhibit a stronger $\tilde m$-dependence compared to models with high
$M_S$. The lower the string scale, the stronger its $\tilde m$-dependence.
However, as can be seen from the right plot of figure \ref{ms}, the $\tilde
m$-dependence becomes almost irrelevant when the string scale is of the order
$10^{16}$ GeV. Finally, cases with $M_S \gtrsim 2\times 10^{17}$ GeV (right part
of the graph) do not satisfy the naturalness criterion (\ref{max}), thus, in the
present context they cannot be considered as viable effective models. We stress
here that in the limit $\tilde m \ra M_Z$ the spectrum becomes fully
supersymmetric for the whole energy range $M_Z-M_S$ and the results of the low
energy supersymmetric models are recovered. We also observe a shift of the string
scale to higher values relative to the non-supersymmetric case discussed in
\cite{Gioutsos:2005uw}.

In summary, in this section we have presented  a string scale classification of
D-brane constructions with $U(3)\times U(2)\times U(1)^N$ ($N=1,2,3$) gauge
symmetry, in the parallel brane scenario and found the following interesting
classes of models with respect to the string scale predictions:%
\vspace{-0.4cm}
\begin{itemize}
  \item[(i)] one class of models with $N=1,3$ abelian branes
             predicts a string scale of the order of the SUSY GUT scale
             i.e.~$M_S \sim 10^{16}$ GeV. Interestingly these models also
             imply $\xi=1$, to a good approximation,
             which means that the non-abelian
             gauge couplings $\alpha_2,\alpha_3$ do unify at $M_S$
  \item[(ii)] in the case of $N=2$ extra abelian branes, and for a specific
              $U(1)$ brane orientation we  find a model with intermediate
              string scale $M_S\sim 10^7-10^8$ GeV
  \item[(iii)] two cases in $N=2,3$ abelian brane scenarios
               predict a low $M_S$ at the $10^4-10^5$ GeV range.
  \item[(iv)] finally, in the $N=2$ abelian case, models $a_2$ allow for a
              string scale of the order of the Planck mass, however, for
              the reasons explained above, these are not considered as
              viable possibilities.
\end{itemize}
\begin{table}[!t]
\renewcommand{\arraystretch}{2.9}
\centering
\begin{tabular}{|c|l|c||c|c|c||c|c|c|c|}
\hline
 & $a_1$ & $b_1$ & $a_2$&$b_2$ &$c_2$ &$a_3$&$b_3$& $c_3$ \\
\hline
 $n_1$ &  $\dsty \frac 23$ & $\dsty  \frac 16$       &  $\dsty \frac 16$
       &  $\dsty \frac 83$ & $\dsty  \frac 23$ & $\dsty  \frac 16$
       &  $\dsty \frac 23$ & $\dsty  \frac 83$ \\
 \hline
 $n_2$ & $\dsty 1$        & $\dsty \frac{\xi'}{2}$ & $\dsty \xi'$
       & $\dsty 1+2\xi'$  & $\dsty 1+2\xi'$
       & $\dsty \frac{3\xi'}{2}$ & $\dsty 1+4\xi'$  & $\dsty 1+2\xi'$ \\
 \hline
\end{tabular}
\caption{\label{kyint}Various $n_{1,2}$ values in the case of intersecting branes
for the models of Table \ref{t1}.}
\end{table}

\subsection{The string scale at intersecting brane scenarios}

We have presented above a classification of the models which arise from various
superpositions of the abelian branes with the non-abelian ones. In these models
the abelian gauge couplings, at the string scale, have equal initial values with
the $\alpha_2$ or $\alpha_3$ couplings. However, in the general case of
intersecting branes the $U(1)$ gauge couplings $\alpha_i'$ are not necessarily
equal to any of the non-abelian gauge couplings. Instead, depending on the
details of the particular D-brane construction, they can take arbitrary values
and as a consequence they imply different predictions for the string scale $M_S$.
In this section, we wish to explore the possible variation of the string scale on
a general perspective, i.e., without imposing specific brane orientations. In
order to facilitate the subsequent analysis, we define a common average value
$\xi'$ for the gauge coupling ratios $\alpha_2/\alpha_i'$ and express $k_Y$ as%
\ba%
 k_Y&\equiv&{6 k_3^2}\,\xi+4 k_2^2+2\xi'\,\sum_{i=1}^N
{k_i'}^2\,\label{kYA}
\ea%
which means that the string scale depends now on two arbitrary parameters, namely
$\xi$ and $\xi'$. In Table~\ref{kyint} we present the values of $n_1$, $n_2$ for
all models of Table \ref{t1}.
\begin{figure}[!t]
\centering
\includegraphics[width=0.49\textwidth]{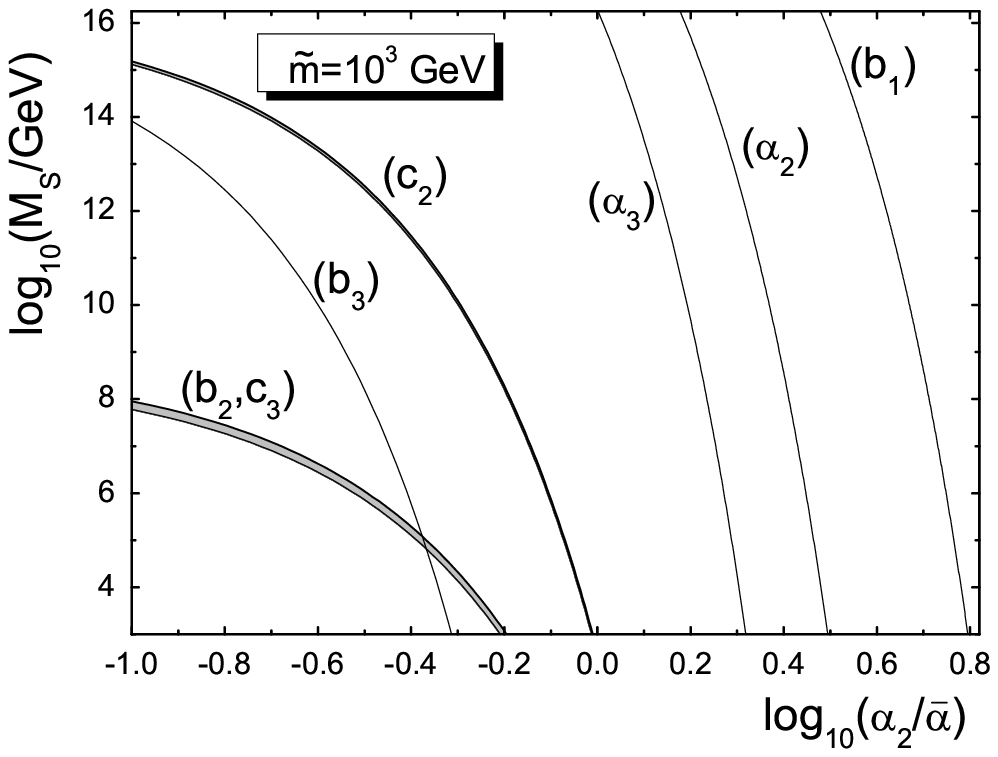}
\includegraphics[width=0.49\textwidth]{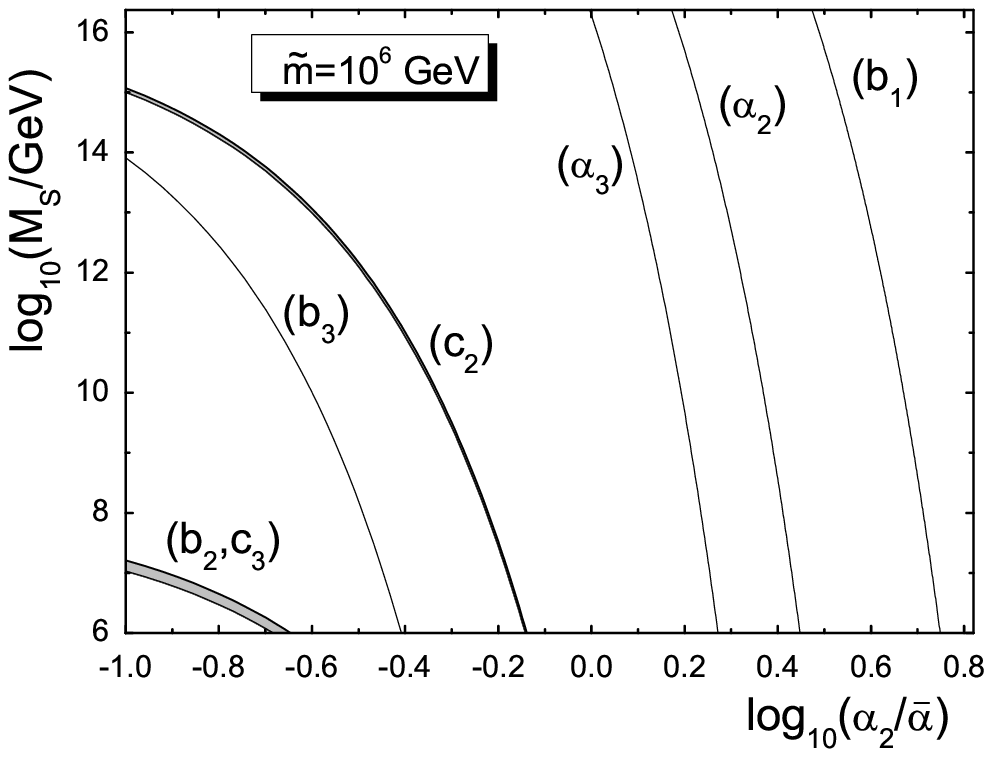}
\\
\includegraphics[width=0.49\textwidth]{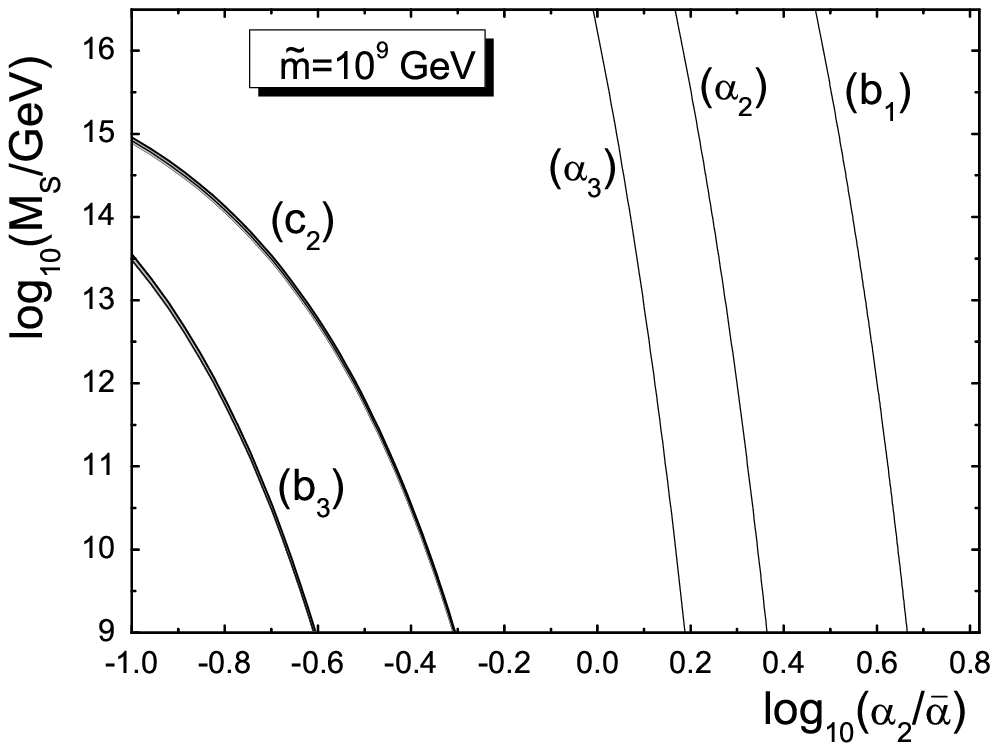}
\includegraphics[width=0.49\textwidth]{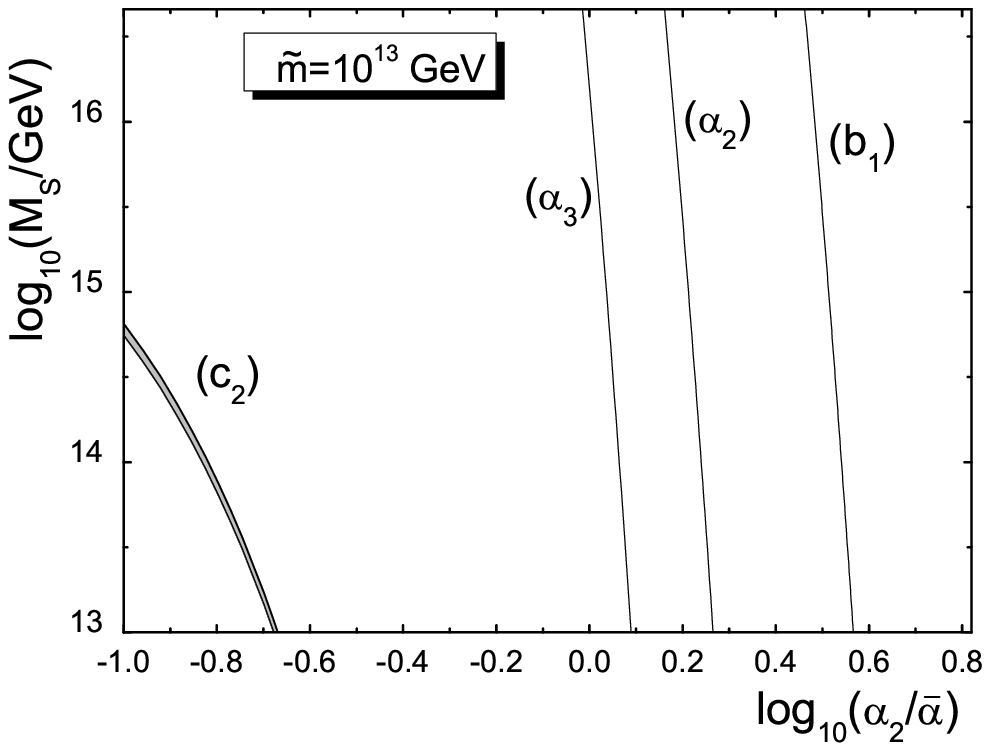}
\caption{\label{msvsxip}The string scale as a function of the parameter $\xi'$
when the breaking of supersymmetry occurs at $\tilde{m}=10^3,10^6,10^9,10^{13}$
GeV. The thickness of the curves corresponds to the $\alpha_3$ experimental
uncertainties at $M_Z$.}
\end{figure}%

Considering the models $a_{2,3}, b_{1,2,3}$ and $c_2$, we have plotted in
Fig.~\ref{msvsxip} the variation of the string scale $M_S$ as a function of the
gauge coupling ratio $\log\xi'=\log(\alpha_2/\bar\alpha)$ ($\bar\alpha$ stands
for the average value of the abelian couplings $\alpha_i'$), using four
characteristic values of the split-susy scale namely $\tilde m = 10^3, 10^6,
10^9, 10^{13}$ GeV.

Starting our discussion with the case $\tilde m=10^3$ GeV (upper left plot of
Fig.~\ref{msvsxip}) we see that in models $b_{2,3}$ and $c_{2,3}$, as $\xi'$
varies the string scale takes values in a large range, however, it is always
lower than the traditional SUSY  GUT scale at least by one order of magnitude.
More precisely, for $\tilde m\approx 10^3$ GeV as we vary $\xi,\xi'$ model $b_2$
gives a string scale in the range $M_S=10^3-10^8$ GeV. In addition, for a
specific value of the gauge coupling ratio $\xi'$ the $b_2$ and $b_3$
$M_S$-curves cross each other, thus they give an identical string mass prediction
around $10^5$ GeV. From Fig.~\ref{msvsxip} we also find that, for any value of
$M_S$ we have $\xi<1$ and $\xi'\le 1$ implying that\footnote{%
The string gauge couplings $\alpha_i'$ are found in the perturbative region.}
$\alpha_2< \alpha_3$ and $\alpha_2\le \bar \alpha$. Of particular interest for
low energy neutrino physics are the models $b_3$  and $c_2$ where $M_S$ raises up
to $10^{14}$ GeV and $10^{15}$ GeV respectively. Indeed, in these cases the mass
of the right-handed neutrino can be of the same order with $M_S$, allowing thus a
see-saw type Majorana mass for its left-handed component in the sub-eV range,
consistent with the experimental value and the present cosmological bounds.

In a second class of models, namely $a_2,a_3,b_1$, the string scale-curves are
steep thus, $M_S$ is more sensitive to the variation of the ratio $\xi'$ as
compared to models discussed previously. As we move to higher scales $\tilde m$,
the $M_S$ curves become  steeper. Moreover, in contrast to the previous cases,
$M_S$ can attain large values of the order $\sim 10^{16}$ GeV or higher. We also
find that $\alpha_2 \ge \bar\alpha$, while for $M_S$ of the order of the SUSY GUT
scale and specific $\xi'$-values, which depend on the particular model, we may
have complete unification ($\xi=1$) of the non-abelian couplings $\alpha_{2,3}$
at $M_S$. In all cases, the highest $M_S$ values (represented by the upper end
points of the curves) correspond to the case of equal non-abelian couplings,
i.e.~$\alpha_3(M_S)=\alpha_2(M_S)$. Besides, for $\tilde m=10^{3}$ GeV we find
that $\xi'\sim 1,1.5,3$ for the models $a_3, a_2, b_1$ respectively, while the
highest unification $M_S$ values  are in the range $[1.75-3.45]\times 10^{16}$
GeV. For $\tilde m =10^{13}$ GeV, $M_S$ raises up to $[4.5-9.0]\times 10^{16}$
GeV.

Similar conclusions can be extracted from the remaining two plots of the same
figure which correspond to the cases of  $\tilde m=10^6$, $10^9$ GeV.  Note that
in the third plot where $\tilde m=10^9$ GeV, the $b_2$-curve does not exist since
in this model the string scale cannot be higher than $\sim 10^8$ GeV. In the
fourth plot ($\tilde m\sim 10^{13}$ GeV), the $b_2,c_3$ and $b_3$ $M_S$-plots are
not present for the same reason. We finally note that models $a_2, a_3, b_1$, for
appropriate gauge coupling values, can also accommodate a right-handed neutrino
at the required mass scale $M_{\nu^c}\sim M_S\ge 10^{14}$ GeV.

\section{ b-$\tau$ unification }

One of the most interesting features in traditional grand unified theories, is
the relations they imply for the third generation Yukawa couplings. In
particular, for a wide class of SUSY GUT models, the equality of the bottom --
tau Yukawa couplings has been shown to be in accordance with the low energy $m_b
/ m_{\tau}$ mass ratio. In this section, in order to see whether the present
D-brane inspired models share this property, we use a renormalization group
approach to examine the $b-\tau$ Yukawa coupling relation at the string scale.

Our procedure is the following. Using the experimentally determined values for
the bottom, tau fermion masses $( m_b, m_\tau )$ we run the 2-loop $SU(3)_C
\times U(1)_Y$ RGE system~\cite{Gioutsos:2005uw,Arason:1991ic} up to the weak
scale ($M_Z$) and reconcile there the results with the well known experimental
values of the weak mixing angle and the gauge couplings. For the scales above
$M_Z$ we consider a split supersymmetric
theory~\cite{Arkani-Hamed:2004fb,Arkani-Hamed:2004yi} where supersymmetry is
broken at an energy scale $\tilde{m}$ generally far above the TeV scale. This
means that the theory above $\tilde{m}$ is fully supersymmetric and the RGE
system involves $\alpha_3, \alpha_2, \alpha_Y, Y_t, Y_b, Y_{\tau}$ couplings (see
Appendix) while below $\tilde{m}$ the effective theory is obtained by removing
squarks, sleptons, charged and pseudoscalar Higgs from the supersymmetric
standard model. The spectrum of Split Supersymmetry ($\mu<\tilde{m}$) contains
the Higgsino components (${\tilde H}_{u,d}$), the gluino ($\tilde g$), the wino
($\tilde W$), the bino ($\tilde B$) and the SM particles with one Higgs doublet
$H$. The relevant 1-loop RGE system involves $\alpha_3, \alpha_2, \alpha_Y, Y'_t,
Y'_b, Y'_{\tau}, \alpha_u, \alpha'_u, \alpha_d, \alpha'_d$ where $ Y'_{t,b,\tau}$
are the Yukawa couplings and $\alpha_{u,d}, \alpha_{u,d}'$ are the gaugino
couplings (relevant equations are summarized in Appendix).

Once the Higgs doublet $H$ is fine tuned to have small mass term, the matching
conditions of the coupling constants at the scale
$\tilde{m}$ become%
\ba%
Y'_{t} = Y_t \sin^2\beta         \qquad\qquad\qquad
                 \alpha'_u = \alpha_Y \sin^2\beta \nonumber\\
Y'_{b} = Y_b \cos^2\beta         \qquad\qquad\qquad
                 \alpha_u = \alpha_2 \sin^2\beta \nonumber \\
Y'_{\tau} = Y_{\tau} \cos^2\beta \qquad\qquad\qquad
                 \alpha'_d = \alpha_Y \cos^2\beta\nonumber \\
                 \alpha_d = \alpha_2 \cos^2\beta \nonumber
\ea%
where the angle $\beta$ is related to the vevs of the SUSY Higgs doublets by
$\tan\beta= \upsilon_u / \upsilon_d$.

Let us briefly describe our strategy to solve the RGEs. In order to determine the
string scale $M_S$, using the experimental values for $\alpha_3, \alpha_e,
\sin^2\theta_W$, we evolve the gauge coupling RGEs above the electroweak scale
until the condition%
\beq%
\left[ n_1 \frac{\alpha_2(\mu)}{\alpha_3(\mu)} + n_2 -
\frac{\alpha_2(\mu)}{\alpha_Y(\mu)} \right]_{\mu=M_S} = 0
\eeq%
is satisfied. Our main objective, however, is the evaluation of the quark masses
and the corresponding Yukawa couplings at $M_S$. For this reason we follow a top
-- bottom approach. The required quantities $m^S_{t,b,\tau} \equiv m_{t,b,\tau}
(M_S)$, entering the RGE system as initial conditions, are considered to be
unknown parameters. The RGE system is then evolved down to the scale $\tilde{m}$
where the matching conditions for the Yukawa and gaugino couplings are applied.
We then continue the RGE evolution down to the electroweak scale and determine
the unknown quark masses at $M_S$ by solving numerically the following system of
algebraic equations%
\ba%
\label{algsys}%
\left( v \sqrt{4 \pi Y'_t(\mu)} - \frac{M_t}{1 + \frac{4\alpha_3(\mu)}{3\pi} -
\frac{Y'_t(\mu)}{2 \pi}} \right)_{\mu=M_t} &=& 0 \nonumber\\
m_b(M_Z) - v \sqrt{4 \pi Y'_b(M_Z)}  &=& 0  \\
m_\tau(M_Z) -  v \sqrt{4 \pi Y'_{\tau}(M_Z)} &=& 0 \nonumber
\ea%
where all $Y'$ quantities\footnote{%
When $\tilde{m} < M_t$ we should replace $Y' \rightarrow Y$ in
eq.~(\ref{algsys}). } %
have an intrinsic dependence on $m^S_{t,b,\tau}$ while $v \approx 173.46$ GeV is
the VEV of the Higgs field $H$ related to the Z-boson mass by $M_Z =  v \sqrt{2
\pi(\alpha_Y + \alpha_2)}$. The experimental values we have used for the running
quark masses are $m_{\tau}(m_{\tau}) = 1.777$ GeV, $m_b(m_b) = 4.25 \pm 0.15$ GeV
while the top quark pole mass is $M_t = 172.7 \pm 2.9\,$ GeV \cite{CDFD0lastmt}.

\begin{figure}[!t]
\centering
\includegraphics[width=0.49\textwidth]{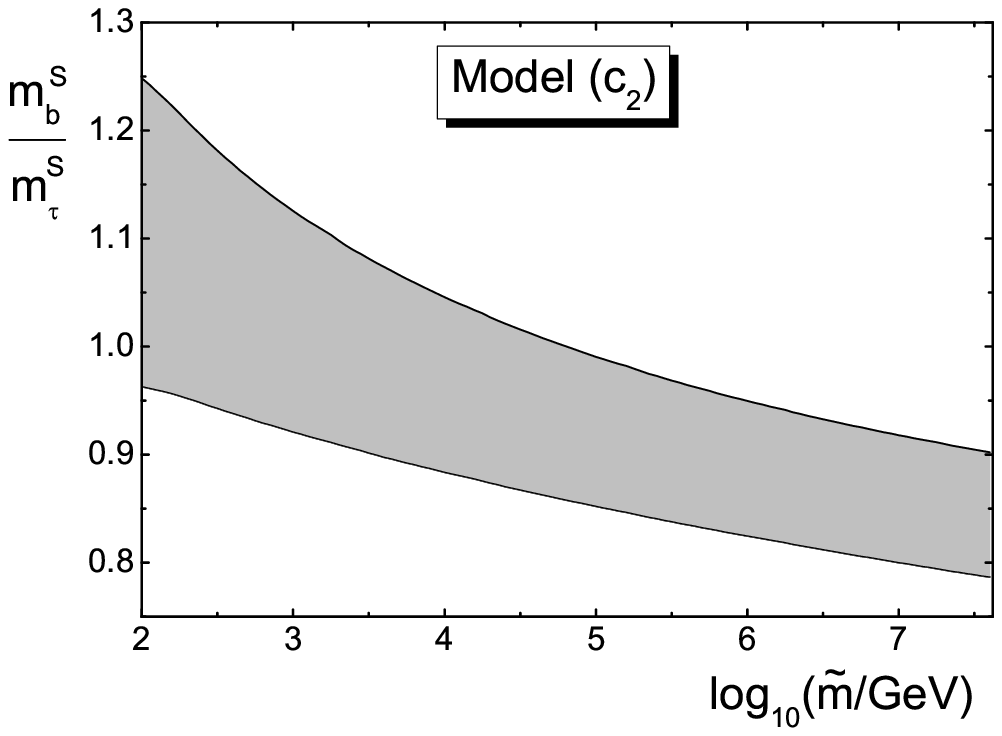}
\includegraphics[width=0.49\textwidth]{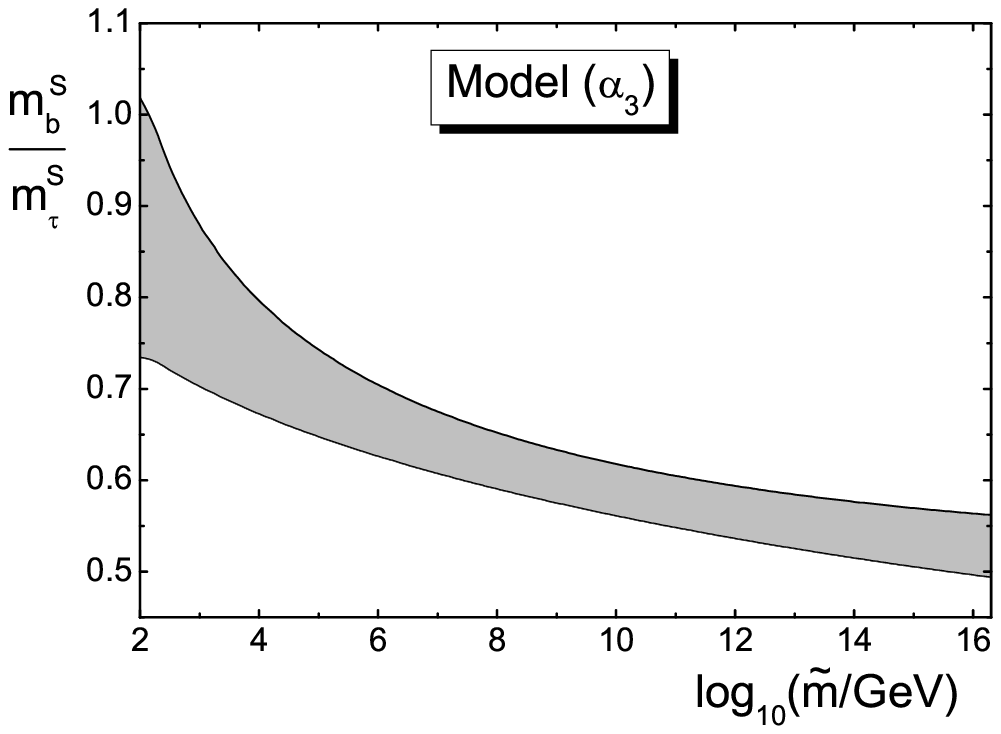}
\\
\includegraphics[width=0.49\textwidth]{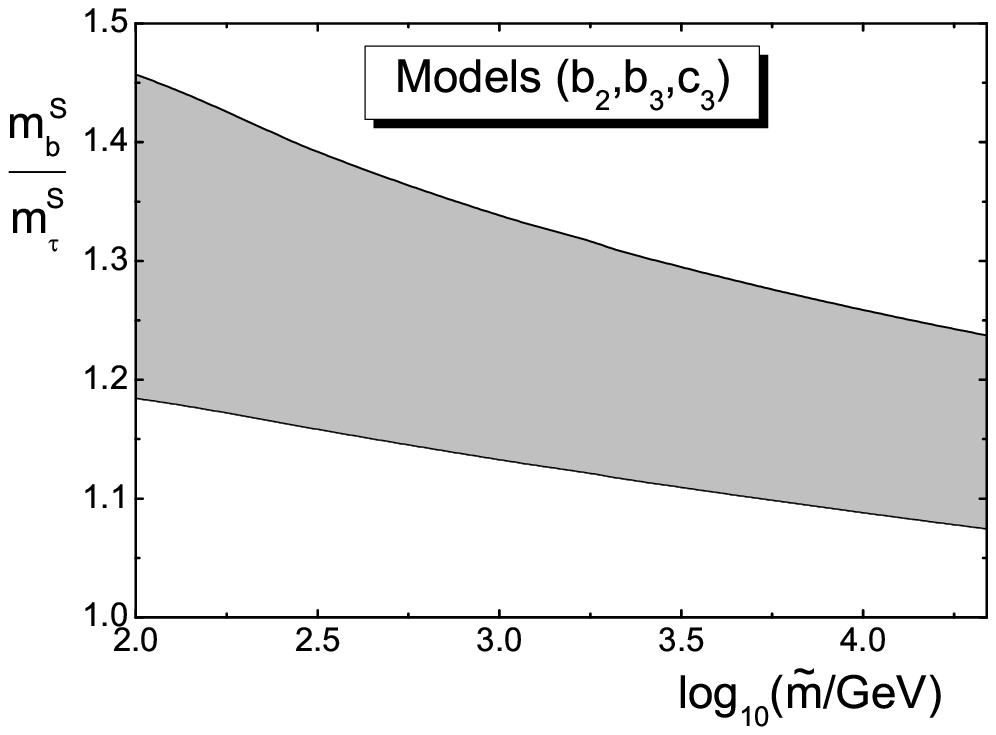}
\includegraphics[width=0.49\textwidth]{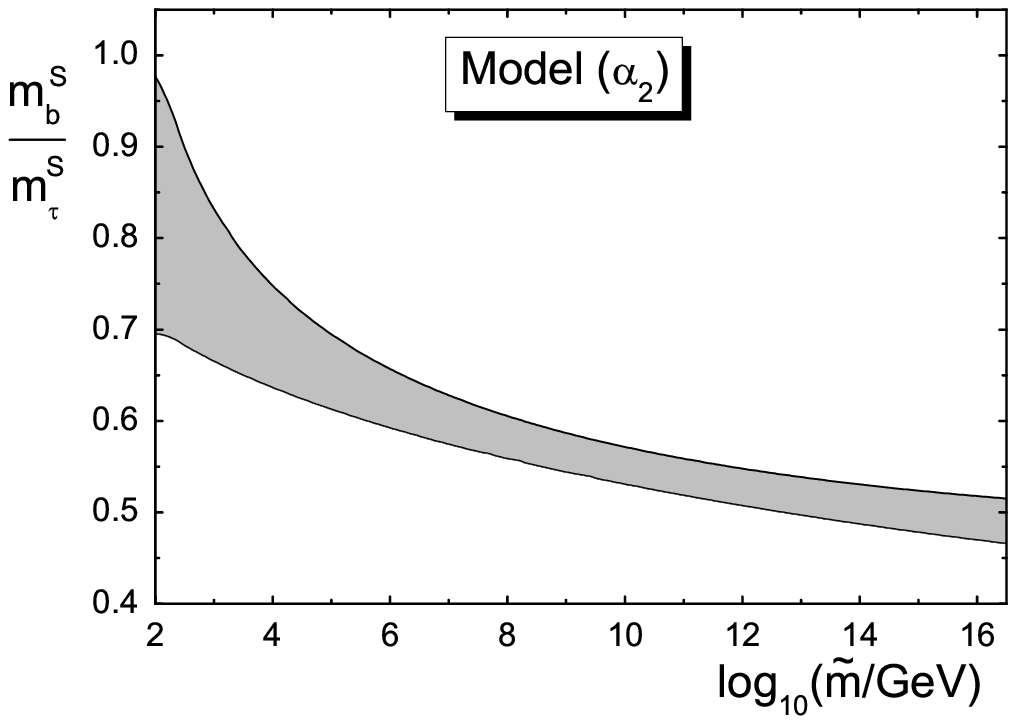}
\caption{\label{mbmtaumoda2} The ratio $m_b/m_\tau$ as a function of the
supersymmetry breaking scale for models $(c_2)$, $(\alpha_3)$, $(b_2,b_3,c_3)$
and $(\alpha_2)$. The shaded region is due to uncertainties in $a_3(M_Z)$,
$m_b(m_b)$ as well as $M_t$.}
\end{figure}%
In Fig.~\ref{mbmtaumoda2} we present our results for $b-\tau$ Yukawa unification
versus the split supersymmetry scale for four characteristics classes of models
in the parallel brane scenario discussed in previous sections. The lower curve of
the shaded bands of the plots corresponds to $\tan\beta = 8.5$ and the upper one
to $\tan\beta = 50$, while $\alpha_3$ uncertainties are also included. In the
right half of the same figure we show the $b-\tau$ relation at $M_S \sim {\cal
O}(10^{16})$ GeV for the models $a_2, a_3$. We observe in these models that the
string scale ratio $m_b / m_{\tau}$ is significantly lower than unity for a wide
range of $\tilde m$, except for low $\tilde m \sim {\cal O}( M_Z)$ values
(i.e.~when theory becomes supesymmetric), where it approaches unity. In the left
half, we see that models $b_{2,3}, c_3$ predict a $b-\tau$ ratio larger than
unity for any value of $\tilde m$. The most interesting result comes from the
model $c_2$, where for a wide range of $\tilde m \sim [10^2-10^6]$ GeV, the
$b-\tau$ ratio (at $M_S \sim 10^{7}$ GeV), is of order one. This result should be
compared with the corresponding non-supersymmetric case where $b-\tau$ Yukawa
equality at $M_S$ holds for a string scale $M_S\sim 2\times 10^6$ GeV.

\section{ Gaugino Masses and the lifetime of the gluino}

In split supersymmetry, gaugino and higgsino masses, unlike their scalar
superpartners, are protected by an R-symmetry and a PQ-symmetry, so that they are
massless at tree-level. It is usually assumed that supersymmetry breaking occurs
in the gravity sector, thus, R-symmetry is broken and gauginos obtain a light
mass by some of the mechanisms described in
refs.~\cite{Arkani-Hamed:2004fb,Antoniadis:2004dt}. At the same time, the
PQ-symmetry, which protects higgsinos from being heavy, must be broken to
generate the $\mu B$ Higgs mixing term. This mass term together with the soft
Higgs doublet masses are appropriately fine-tuned so that one linear Higgs
combination is  light, while the second one is of the order of the split susy
scale $\sim\tilde m^2$. Thus, in this scenario all scalars, but one Higgs linear
combination, are heavy with masses of the order $\tilde m$. This is an important
advantage of split susy, since heavy scalars suppress remarkably the problematic
flavor violating processes of the corresponding SUSY models.

\begin{figure}[!t]
\centering%
    \includegraphics[width=0.49\textwidth]{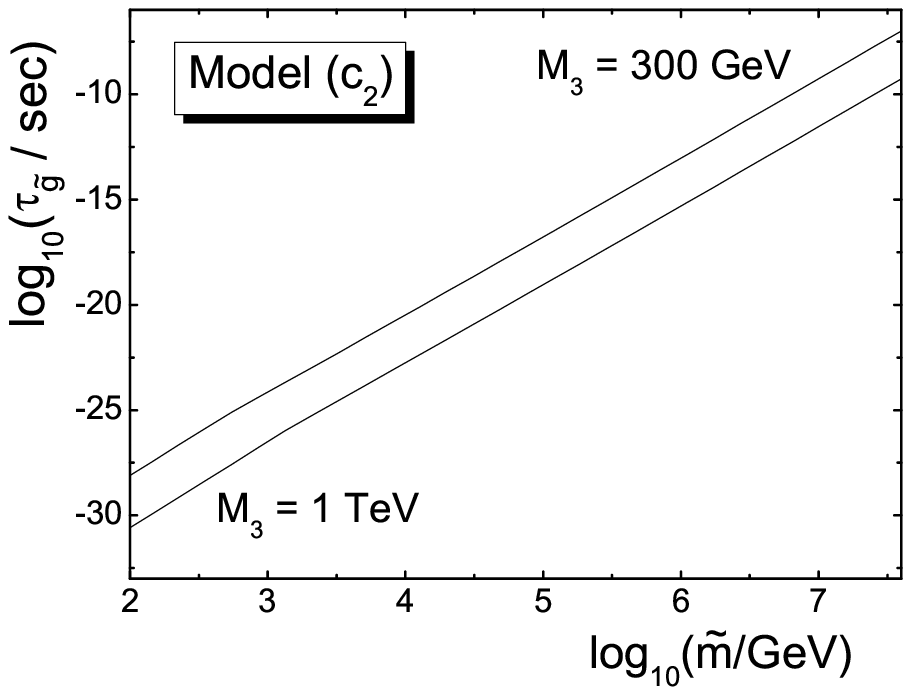}
    \includegraphics[width=0.49\textwidth]{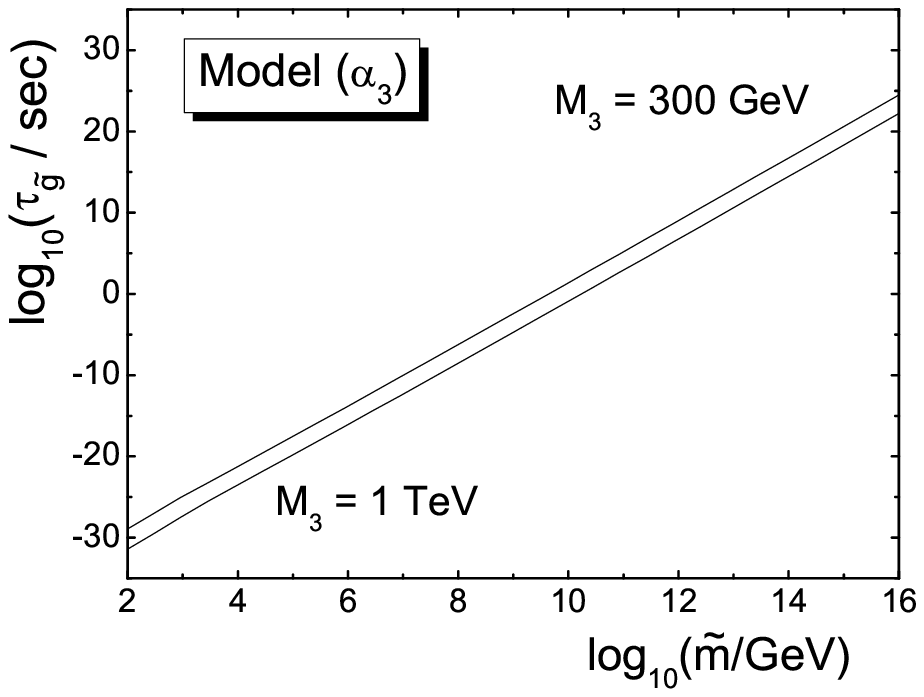}
    \includegraphics[width=0.49\textwidth]{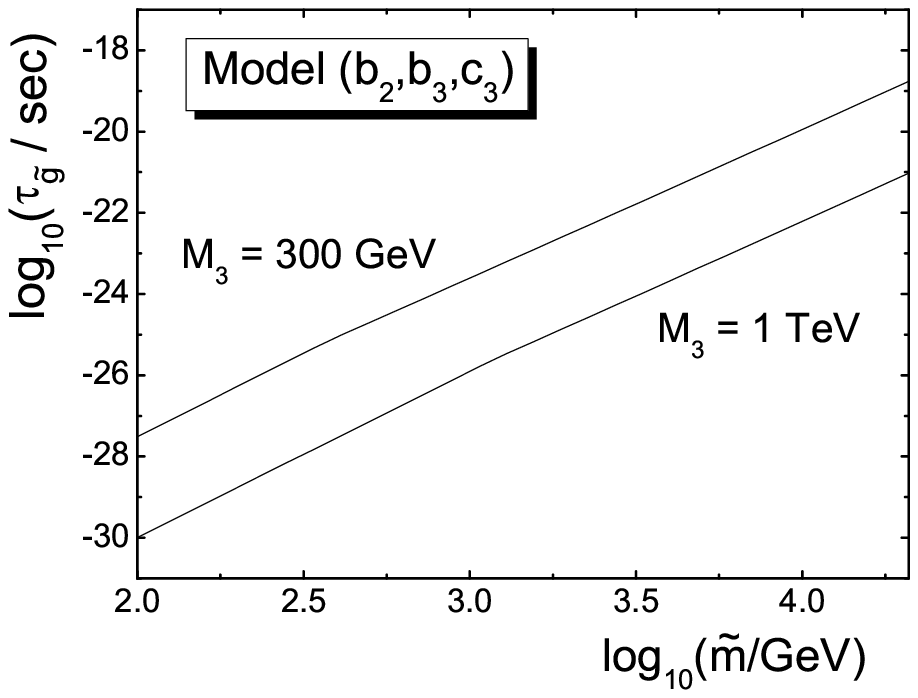}
    \includegraphics[width=0.49\textwidth]{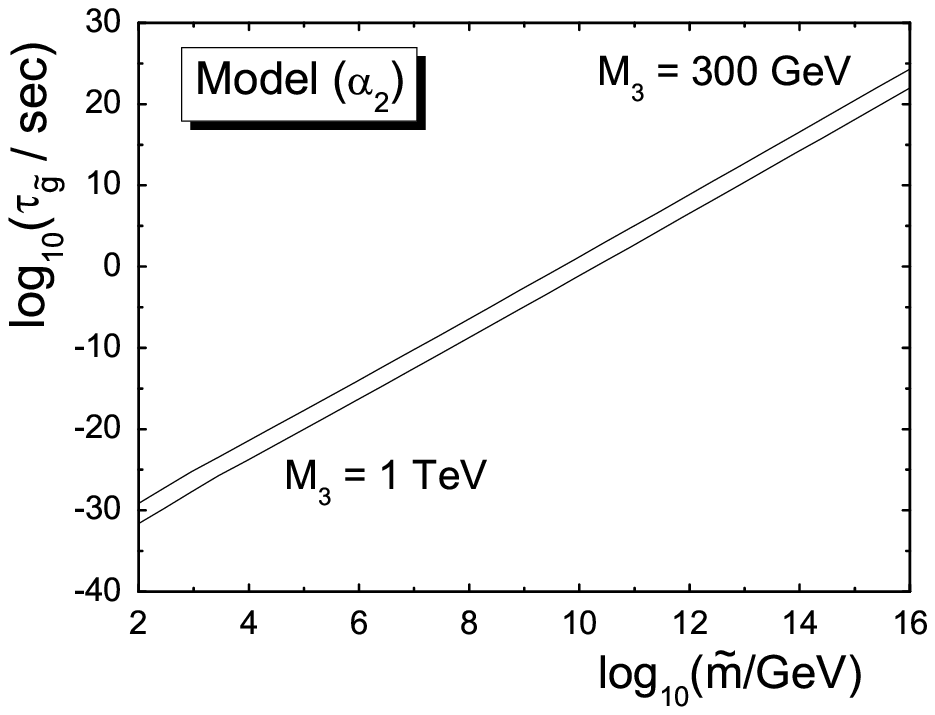}
\caption{\label{glu} The gluino decay life time $\tau_{\tilde{g}}$ in four
distinct cases for $M_3(M_S)=0.3$ and $1$ TeV.}
\end{figure}%
The most distinct signature for a high  scale split supersymmetry is a
slow-decaying gluino. This important indication differentiates the various split
supersymmetry models discussed so far in this work. The lifetime of the gluino,
which is mediated by virtual squark exchange, is given
by~\cite{Antoniadis:2004dt,Gambino:2005eh}%
\ba%
\tau_{\tilde g} &=& 3\times 10^{-2} \left(\frac{\tilde m_{sq}}{10^9{\rm
GeV}}\right)^4\,\left(\frac{1\,{\rm TeV}}{m_{\tilde g}}\right)^5\,{\rm
sec}\label{life}%
\ea%
where $\tilde m_{sq}$ is the squark mass of the order of split-susy scale $\tilde
m$ and $m_{\tilde g}$ is the TeV-scale gluino mass. The $\tau_{\tilde g}$
expression in eq.~(\ref{life}) depends on $\tilde m_{sq}$ and $m_{\tilde g}$
which are expressed in terms of $\tilde m$, thus, the experimental detection of a
long-lived late-decaying gluino would determine the effective split supersymmetry
scale. For  small $\tilde m$ (i.e., less than a few hundreds  TeV), gluinos will
decay within the detectors, however, if the split SUSY scale is sufficiently
large the gluinos can live long enough to travel distances longer than the size
of the detector. In this case, they can form R-handrons that lose energy through
ionization while a fraction of them will stop within the
detector~\cite{Arvanitaki:2005nq}. Several phenomenological studies explored the
possibility of observing interesting signatures at
LHC~\cite{Gambino:2005eh,Cheung:2004ad}.

Using the relevant equations given in the Appendix, we calculate the gluino pole
mass and its lifetime as a function of $\tilde m$, assuming two starting values
for the gaugino mass,  $M_3=300$ GeV and $M_3=1$ TeV. In Fig.~\ref{glu} we plot
the lifetime of the gluino vs the split supersymmetry scale for the four distinct
cases discussed in previous sections. In low string scale models $b_2,b_3, c_3$
the gluino lifetime is  less than $ 10^{-18}$ sec, thus the gluinos will decay
into the detectors. In model $c_2$, for $\tilde m\ge  {\cal O}(10^6)$ GeV, one
obtains $\tau_{\tilde g}\ge 10^{-12}$sec, meaning that a vertex displacement can
be observed in LHC and experimental measurements can determine the value of split
susy scale.

Models $a_2, a_3$ predict  a higher string scale ($ M_S \sim 10^{16}$ GeV or so),
allowing thus for the possibility of high $\tilde m$. Then, the squark mass,
being of the order of the split-susy scale, i.e., $\tilde m_{sq}\sim M_S$,
enhances dramatically the lifetime of the gluino.  As can be seen from the
corresponding plots in Fig.~\ref{glu}, a squark mass in the range $10^{13}$ to
$10^{16}$ GeV, implies\footnote{%
For an incomplete list of works discussing other interesting implications of
split susy, see~\cite{other}.} %
a cosmologically stable gluino with a lifetime in the
range of $10^{10}$ to $10^{26}$ sec. We should also note that there is a
significant dependence of our $M_3$ and $\tau_{\tilde g}$ results, on the initial
value  of the gaugino mass $M_3$ at the string scale $M_S$. This  dependence
becomes more important for large $\tilde m$ values which shows that the
renormalization effects below $\tilde m$ are substantial.

\section{Conclusions}

Inspired by  D-brane models, in the present work we studied extensions of the
Standard Model gauge symmetry of the form $U(3)\times U(2)\times U(1)^N$, in the
context of split supersymmetry. We considered  configurations with one, two and
three ($N=1,2,3$) abelian branes, and made a complete classification of all
models with regard to the various hypercharge embeddings which imply a realistic
particle content.

We started our analysis with the implications of split supersymmetry  on the
string scale, considering first models which arise in parallel brane scenarios
where the $U(1)$ branes are superposed with the $U(2)$ or $U(3)$ brane stacks.
Varying the split susy scale in a wide range, we examined the evolution of the
gauge couplings in the above context and found three distinct classes of models
with the following characteristics: i) one class of models which arises in
configurations with $N=1$ and $N=3$ abelian branes, predicts a string scale of
the order of the SUSY GUT scale $M_S\sim 10^{16}$ GeV; interestingly, these
models also imply that the non-abelian gauge couplings unify
($\alpha_2=\alpha_3$) at $M_S$; ii) in a particular case of $N=2$ abelian branes,
corresponding to a specific $U(1)$ brane orientation we find a model with
intermediate string scale $M_S\sim 10^7-10^8$ GeV, and iii) two cases in $N=2$
and $N=3$ abelian brane scenarios result to a low $M_S$ at the TeV range.
Moreover, we  analyzed the third family fermion mass relations and found that in
the intermediate string scale ($M_S\sim 10^8$GeV) model the low energy ratio
$m_b/m_{\tau}$ is compatible with $b-\tau$ Yukawa unification at the string
scale, for a wide range of split supersymmetry scale $\tilde{m} \sim [0.5 -
10^3]$ TeV.

We further performed a similar analysis considering arbitrary $U(1)_i$-gauge
coupling relations corresponding to possible intersecting brane scenarios and
classified the various models according to their predictions for the magnitude of
the string scale, the gaugino masses and other low energy implications. We found
that the main features observed in the case of parallel brane scenario persist,
however, once we relax the conditions $\alpha_i'=\alpha_{2}$ and
$\alpha_i'=\alpha_{3}$, models previously ruled out due to unacceptably large
string scale $(M_S \ge M_{Pl}$), are now compatible with lower viable string
scales $M_S\le 10^{17}$ GeV for specific $\alpha_i'-\alpha_{2,3}$ gauge coupling
relations at $M_S$. The D-brane configurations proposed here, under the specific
charge assignments are also capable of accommodating  a right-handed neutrino
$\nu^c$. In several viable cases, the string scale is found of the order $M_S \ge
10^{14}$ GeV,  thus a $\nu^c$ mass of the same order arises so that a see-saw
type light left-handed neutrino component is obtained in the sub-eV range as
required by experimental and cosmological data. Finally, we calculated  the
gaugino masses and found how the lifetime of the gluino discriminates the various
models discussed in this work.

{\bf Ackmowledgements}. {\it This research was funded by the programs
`PYTHAGORAS' and `HERAKLEITOS' of the Operational Program for Education and
Initial Vocational Training of the Hellenic Ministry of Education under the 3rd
Community Support Framework and the European Social Fund. G.K.L.~wishes to thank
I.~Antoniadis for discussions.}

\newpage

\section*{Appendix}

In this appendix we collect the renormalization group equations for the split
supersymmetry at one loop level that were used in the analysis of b-$\tau$
unification. The two loop equations can be found in~\cite{Giudice:2004tc}, while
a similar notation but with different conventions can be seen
in~\cite{Huitu:2005ef}. In particular, the 1-loop RGEs for the gauge couplings
are
\ba%
\frac{d \tilde{\alpha}_i}{dt} =  b_i \tilde{\alpha}_i^2
\ea%
where $i=Y,2,3$ and $\tilde\alpha = g^2 / 16 \pi^2$. Below the scale $\tilde{m}$
where supersymmetry is broken the beta coefficients are $(b_Y,b_2,b_3 ) =
(\frac{15}{2}, -\frac 76 , -5)$, while above are $(b_Y^{\rm{su}}, b_2^{\rm{su}},
b_3^{\rm{su}})= (11, 1, -3)$.

Below $\tilde{m}$ the equations that governs the Yukawa ($h$) and gaugino
couplings ($\tilde g$) are also needed. For the Yukawa couplings we have%
\ba%
{d \over dt}\ln\tilde{Y}'_t &=& -8 {\tilde \alpha_3} - {9 \over 4} {\tilde
\alpha_2} - {17 \over 4} {\tilde \alpha_Y} + \frac 92 \tilde{Y}^\prime_t+ \frac
32 \tilde{Y}^\prime_b+ \tilde{Y}^\prime_\tau+ \frac 32{\tilde \alpha_u}+\frac
32{\tilde \alpha_d} + \frac 12 {\tilde \alpha^\prime_u}+ \frac 12{\tilde
\alpha^\prime_d} \qquad
\label{eq-a.13}\\
{d \over dt}\ln\tilde{Y}'_b &=& -8  {\tilde \alpha_3} - {9 \over 4} {\tilde
\alpha_2} - {5 \over 4} {\tilde \alpha_Y} + \frac 32 \tilde{Y}^\prime_t+ \frac 92
\tilde{Y}^\prime_b+ \tilde{Y}^\prime_\tau+ \frac 32 {\tilde \alpha_u} + \frac 32
{\tilde \alpha_d} + \frac 12 {\tilde \alpha^\prime_u}+ \frac 12
{\tilde \alpha^\prime_d} \\
{d \over dt}\ln\tilde{Y}'_\tau &=& -{9 \over 4} {\tilde \alpha_2} - {15 \over 4}
{\tilde \alpha_Y} + 3\tilde{Y}^\prime_t+3\tilde{Y}^\prime_b+\frac 52
\tilde{Y}^\prime_\tau+ \frac 32 {\tilde \alpha_u}+ \frac 32 {\tilde \alpha_d}
+\frac 12 {\tilde \alpha^\prime_u}+ \frac 12 {\tilde \alpha^\prime_d}
\ea%
where $\tilde{Y}' = h^2 / 16\pi^2$ and $\tilde\alpha = {\tilde{g}^2 /16\pi^2}$.
For the gaugino couplings the equations are%
\ba%
{d{\tilde \alpha_u} \over dt} &=& - 3 {\tilde \alpha_u} \left({11 \over 4}
{\tilde \alpha_2} + {1 \over 4} {\tilde \alpha_Y}\right) + {1 \over 4} {\tilde
\alpha_u}(5{\tilde \alpha_u}-2{\tilde \alpha_d}+{\tilde \alpha^\prime_u})
+{({\tilde \alpha_u}{\tilde \alpha_d}{\tilde \alpha^\prime_u}{\tilde
\alpha^\prime_d})}^{1/2} \nonumber\\
& & +{1 \over 2}{\tilde
\alpha_u}(6\tilde{Y}^\prime_t+6\tilde{Y}^\prime_b+2\tilde{Y}^\prime_\tau+
3{\tilde \alpha_u}+3{\tilde \alpha_d}+{\tilde \alpha^\prime_u}+{\tilde
\alpha^\prime_d})
\\[1.2ex]
 {d{\tilde \alpha^\prime_u} \over dt} &=& -3{\tilde
\alpha^\prime_u} \left({3 \over 4} {\tilde \alpha_2} + {1 \over 4} {\tilde
\alpha_Y}\right) + {3 \over 4} {\tilde \alpha^\prime_u}({\tilde
\alpha^\prime_u}+2{\tilde \alpha^\prime_d}+ {\tilde \alpha_u})+3{({\tilde
\alpha_u}{\tilde \alpha_d}{\tilde \alpha^\prime_u}{\tilde
\alpha^\prime_d})}^{1/2} \nonumber
\\
&&+{1 \over 2}{\tilde
\alpha^\prime_u}(6\tilde{Y}^\prime_t+6\tilde{Y}^\prime_b+2\tilde{Y}^\prime_\tau+
3{\tilde \alpha_u}+3{\tilde \alpha_d}+{\tilde \alpha^\prime_u}+{\tilde
\alpha^\prime_d})
\\[1.2ex]
{d{\tilde \alpha_d} \over dt} &=& -3 {\tilde \alpha_d} \left({11 \over 4} {\tilde
\alpha_2} + {1 \over 4} {\tilde \alpha_Y}\right) + {1 \over 4} {\tilde
\alpha_d}(-2{\tilde \alpha_u}+5{\tilde \alpha_d}+{\tilde \alpha^\prime_d})
+{({\tilde \alpha_u}{\tilde \alpha_d}{\tilde \alpha^\prime_u}{\tilde
\alpha^\prime_d})}^{1/2} \nonumber
\\
&&+{1 \over 2}{\tilde
\alpha_d}(6\tilde{Y}^\prime_t+6\tilde{Y}^\prime_b+2\tilde{Y}^\prime_\tau+
3{\tilde \alpha_u}+3{\tilde \alpha_d}+{\tilde \alpha^\prime_u}+{\tilde
\alpha^\prime_d})
\\[1.2ex]
  {d{\tilde \alpha^\prime_d} \over dt} &=& -3{\tilde
\alpha^\prime_d} \left({3 \over 4} {\tilde \alpha_2} + {1 \over 4} {\tilde
\alpha_Y}\right) + {3 \over 4} {\tilde \alpha^\prime_d}({\tilde
\alpha^\prime_d}+2{\tilde \alpha^\prime_u}+ {\tilde \alpha_d})+3{({\tilde
\alpha_u}{\tilde \alpha_d}{\tilde \alpha^\prime_u}{\tilde
\alpha^\prime_d})}^{1/2} \nonumber
\\
&&+{1 \over 2}{\tilde
\alpha^\prime_d}(6\tilde{Y}^\prime_t+6\tilde{Y}^\prime_b+2\tilde{Y}^\prime_\tau+
3{\tilde \alpha_u}+3{\tilde \alpha_d}+{\tilde \alpha^\prime_u}+{\tilde
\alpha^\prime_d})%
\ea%
Moving above $\tilde{m}$ up to the string scale $M_S$, the equations that
describe the evolution of the Yukawa couplings ($\lambda$) are
\begin{eqnarray}
  \frac{d}{dt}\ln\tilde Y_t&=&  -\frac{13}{9}\tila_Y
  - 3 \tila_2 - \frac{16}{3} \tila_3 + 6 \tilde Y_t +  \tilde Y_b   \nonumber\\
 \frac{d}{dt}\ln\tilde Y_b &=&-\frac{7}{9} \tila_Y
  - 3 \tila_2 -\frac{16}{3} \tila_3 +  \tilde Y_t + 6 \tilde Y_b + \tilde Y_{\tau} \\
   \frac{d}{dt}\ln\tilde Y_\tau &=&  -3 \tila_Y
  - 3 \tila_2 + 3 \tilde Y_b +4 \tilde Y_{\tau}    \nonumber
\end{eqnarray}
where $\tilde{Y} = \lambda^2 / 16\pi^2$. The RGE for the $M_3$ gaugino mass at
one loop level below the split SUSY scale is
 \ba
{d \over dt} \ln M_3 &=&  - 9{\tilde \alpha_3} (1+c_{\tilde g}{\tilde \alpha_3}),
\ea%
where $c_{\tilde g}$ = 38/3 in ${\overline {MS}}$ scheme and $c_{\tilde g}= 10$
in ${\overline {DR}}$. Above $\tilde m$ the previous equation becomes%
\ba%
{d \over dt} \ln M_3 &=& b_3^{SU} {\tilde \alpha_3}
\ea%
\newpage

\end{document}